\documentclass[12pt,english]{article}
\usepackage{mathptmx}
\usepackage[T1]{fontenc}
\usepackage[latin9]{inputenc}
\usepackage{geometry}
\usepackage{color}
\usepackage{babel}
\usepackage{mathtools}
\usepackage{amsmath}
\usepackage{amssymb}
\usepackage{stmaryrd}
\usepackage{graphicx}
\usepackage{setspace}
\usepackage[authoryear,semicolon]{natbib}
\makeatletter
\@ifundefined{date}{}{\date{}}
\usepackage{babel}

\usepackage{times}
\usepackage{pdfsync}
\usepackage[font={small}]{caption}

\bibpunct{(}{)}{,}{a}{,}{,}

\usepackage{amsmath}
\usepackage{subfig}
\usepackage{floatrow}

\makeatother

\begin{document}
\global\long\def\real{\mathbb{R}}%
 
\global\long\def\RefVol{\Omega_{0}}%
 
\global\long\def\Refx{\mathbf{X}}%
 
\global\long\def\Curx{\mathbf{x}}%
 
\global\long\def\map{\boldsymbol{\chi}}%
 
\global\long\def\defgrad{\mathbf{F}}%
 
\global\long\def\defgradT{\mathbf{F}^{\mathrm{T}}}%
 
\global\long\def\defgradTi{\mathbf{F}^{\mathrm{-T}}}%
 
\global\long\def\d{\mathrm{d}}%
 
\global\long\def\RCG{\mathbf{C}}%
 
\global\long\def\LCG{\mathbf{b}}%
 
\global\long\def\div#1{\nabla\cdot#1}%
 
\global\long\def\curl#1{\nabla\times#1}%
 
\global\long\def\T#1{#1^{\mathrm{T}}}%
 
\global\long\def\CurStress{\boldsymbol{\sigma}}%
 
\global\long\def\Refcurl#1{\nabla_{\mathbf{X}}\times#1}%
 
\global\long\def\s#1{#1^{\star}}%
 
\global\long\def\m#1{#1^{(m)}}%
 
\global\long\def\f#1{#1^{(f)}}%
 
\global\long\def\p#1{#1^{(p)}}%
 
\global\long\def\Refdiv#1{\nabla_{\mathbf{X}}\cdot#1}%
 
\global\long\def\phase#1{#1^{\left(p\right)}}%
 
\global\long\def\jm{j_{m}^{(p)}}%
 
\global\long\def\xinc{\mathbf{\dot{x}}}%
 
\global\long\def\Pinc{\mathbf{\dot{P}}}%
 
\global\long\def\Einc{\mathbf{\dot{E}}}%
 
\global\long\def\Finc{\mathbf{\dot{F}}}%
 
\global\long\def\dinc{\mathbf{\check{d}}}%
 
\global\long\def\einc{\mathbf{\check{e}}}%
 
\global\long\def\Dinc{\mathbf{\dot{D}}}%
 
\global\long\def\sinc{\boldsymbol{\Sigma}}%
 
\global\long\def\fA{\mathbb{\boldsymbol{\mathscr{A}}}}%
 
\global\long\def\fB{\mathbb{\boldsymbol{\mathscr{B}}}}%
 
\global\long\def\fC{\mathbb{\boldsymbol{\mathscr{C}}}}%
 
\global\long\def\plane{\left(x_{1},x_{3}\right)}%
 
\global\long\def\anti{\dot{x}_{2}}%
 
\global\long\def\G{\mathbf{G}}%
 
\global\long\def\Gt{\mathbf{G}'}%
 
\global\long\def\k{\mathbf{k}}%
 
\global\long\def\Gtk{\left(\Gt+\k\right)}%
 
\global\long\def\Gk{\left(\G+\k\right)}%
 
\global\long\def\GGtk{\left(\G+\Gt+\k\right)}%
 
\global\long\def\ebias{\hat{e}}%
 
\global\long\def\deformation{\boldsymbol{\chi}}%
 
\global\long\def\dg{\mathbf{F}}%
 
\global\long\def\dgcomp#1{F_{#1}}%
 
\global\long\def\piola{\mathbf{P}}%
 
\global\long\def\refbody{\Omega_{0}}%
 
\global\long\def\refbnd{\partial\refbody}%
 
\global\long\def\bnd{\partial\Omega}%
 
\global\long\def\rcg{\mathbf{C}}%
 
\global\long\def\lcg{\mathbf{b}}%
 
\global\long\def\rcgcomp#1{C_{#1}}%
 
\global\long\def\cronck#1{\delta_{#1}}%
 
\global\long\def\lcgcomp#1{b_{#1}}%
 
\global\long\def\deformation{\boldsymbol{\chi}}%
 
\global\long\def\dgt{\dg^{\mathrm{T}}}%
 
\global\long\def\idgcomp#1{F_{#1}^{-1}}%
 
\global\long\def\velocity{\mathbf{v}}%
 
\global\long\def\accel{\mathbf{a}}%
 
\global\long\def\vg{\mathbf{l}}%
 
\global\long\def\idg{\dg^{-1}}%
 
\global\long\def\cauchycomp#1{\sigma_{#1}}%
 
\global\long\def\idgt{\dg^{\mathrm{-T}}}%
 
\global\long\def\cauchy{\boldsymbol{\sigma}}%
 
\global\long\def\normal{\mathbf{n}}%
 
\global\long\def\normall{\mathbf{N}}%
 
\global\long\def\traction{\mathbf{t}}%
 
\global\long\def\tractionl{\mathbf{t}_{L}}%
 
\global\long\def\ed{\mathbf{d}}%
 
\global\long\def\edcomp#1{d_{#1}}%
 
\global\long\def\edl{\mathbf{D}}%
 
\global\long\def\edlcomp#1{D_{#1}}%
 
\global\long\def\ef{\mathbf{e}}%
 
\global\long\def\efcomp#1{e_{#1}}%
 
\global\long\def\efl{\mathbf{E}}%
 
\global\long\def\freech{q_{e}}%
 
\global\long\def\surfacech{w_{e}}%
 
\global\long\def\outer#1{#1^{\star}}%
 
\global\long\def\perm{\epsilon_{0}}%
 
\global\long\def\matper{\epsilon}%
 
\global\long\def\jump#1{\llbracket#1\rrbracket}%
 
\global\long\def\identity{\mathbf{I}}%
 
\global\long\def\area{\mathrm{d}a}%
 
\global\long\def\areal{\mathrm{d}A}%
 
\global\long\def\refsys{\mathbf{X}}%
 
\global\long\def\Grad{\nabla_{\refsys}}%
 
\global\long\def\grad{\nabla}%
 
\global\long\def\divg{\nabla\cdot}%
 
\global\long\def\Div{\nabla_{\refsys}}%
 
\global\long\def\derivative#1#2{\frac{\partial#1}{\partial#2}}%
 
\global\long\def\aef{\Psi}%
 
\global\long\def\dltendl{\edl\otimes\edl}%
 
\global\long\def\ii#1{I_{#1}}%
 
\global\long\def\dh{\hat{D}}%
 
\global\long\def\inc#1{\dot{#1}}%
 
\global\long\def\sys{\mathbf{x}}%
 
\global\long\def\curl{\nabla}%
 
\global\long\def\Curl{\nabla_{\refsys}}%
 
\global\long\def\piolaincpush{\boldsymbol{\Sigma}}%
 
\global\long\def\piolaincpushcomp#1{\Sigma_{#1}}%
 
\global\long\def\edlincpush{\check{\mathbf{d}}}%
 
\global\long\def\edlincpushcomp#1{\check{d}_{#1}}%
 
\global\long\def\efincpush{\check{\mathbf{e}}}%
 
\global\long\def\efincpushcomp#1{\check{e}_{#1}}%
 
\global\long\def\elaspush{\boldsymbol{\mathcal{C}}}%
 
\global\long\def\elecpush{\boldsymbol{\mathcal{A}}}%
 
\global\long\def\elaselecpush{\boldsymbol{\mathcal{B}}}%
 
\global\long\def\disgrad{\mathbf{h}}%
 
\global\long\def\disgradcomp#1{h_{#1}}%
 
\global\long\def\trans#1{#1^{\mathrm{T}}}%
 
\global\long\def\phase#1{#1^{\left(p\right)}}%
 
\global\long\def\elecpushcomp#1{\mathcal{A}_{#1}}%
 
\global\long\def\elaselecpushcomp#1{\mathcal{B}_{#1}}%
 
\global\long\def\elaspushcomp#1{\mathcal{C}_{#1}}%
 
\global\long\def\dnh{\aef_{DH}}%
 
\global\long\def\woo{\varsigma}%
 
\global\long\def\wif{\Lambda}%
 
\global\long\def\structurefun{S}%
 
\global\long\def\dg{\mathbf{F}}%
 
\global\long\def\dgcomp#1{F_{#1}}%
 
\global\long\def\piola{\mathbf{P}}%
 
\global\long\def\refbody{\Omega_{0}}%
 
\global\long\def\refbnd{\partial\refbody}%
 
\global\long\def\bnd{\partial\Omega}%
 
\global\long\def\rcg{\mathbf{C}}%
 
\global\long\def\lcg{\mathbf{b}}%
 
\global\long\def\rcgcomp#1{C_{#1}}%
 
\global\long\def\cronck#1{\delta_{#1}}%
 
\global\long\def\lcgcomp#1{b_{#1}}%
 
\global\long\def\deformation{\boldsymbol{\chi}}%
 
\global\long\def\dgt{\dg^{\mathrm{T}}}%
 
\global\long\def\idgcomp#1{F_{#1}^{-1}}%
 
\global\long\def\velocity{\mathbf{v}}%
 
\global\long\def\accel{\mathbf{a}}%
 
\global\long\def\vg{\mathbf{l}}%
 
\global\long\def\idg{\dg^{-1}}%
 
\global\long\def\cauchycomp#1{\sigma_{#1}}%
 
\global\long\def\idgt{\dg^{\mathrm{-T}}}%
 
\global\long\def\cauchy{\boldsymbol{\sigma}}%
 
\global\long\def\normal{\mathbf{n}}%
 
\global\long\def\normall{\mathbf{N}}%
 
\global\long\def\traction{\mathbf{t}}%
 
\global\long\def\tractionl{\mathbf{t}_{L}}%
 
\global\long\def\ed{\mathbf{d}}%
 
\global\long\def\edcomp#1{d_{#1}}%
 
\global\long\def\edl{\mathbf{D}}%
 
\global\long\def\edlcomp#1{D_{#1}}%
 
\global\long\def\ef{\mathbf{e}}%
 
\global\long\def\efcomp#1{e_{#1}}%
 
\global\long\def\efl{\mathbf{E}}%
 
\global\long\def\freech{q_{e}}%
 
\global\long\def\surfacech{w_{e}}%
 
\global\long\def\outer#1{#1^{\star}}%
 
\global\long\def\perm{\epsilon_{0}}%
 
\global\long\def\matper{\epsilon}%
 
\global\long\def\jump#1{\llbracket#1\rrbracket}%
 
\global\long\def\identity{\mathbf{I}}%
 
\global\long\def\area{\mathrm{d}a}%
 
\global\long\def\areal{\mathrm{d}A}%
 
\global\long\def\refsys{\mathbf{X}}%
 
\global\long\def\Grad{\nabla_{\refsys}}%
 
\global\long\def\grad{\nabla}%
 
\global\long\def\divg{\nabla\cdot}%
 
\global\long\def\Div{\nabla_{\refsys}}%
 
\global\long\def\derivative#1#2{\frac{\partial#1}{\partial#2}}%
 
\global\long\def\aef{\Psi}%
 
\global\long\def\dltendl{\edl\otimes\edl}%
 
\global\long\def\tr#1{\mathrm{tr}\left(#1\right)}%
 
\global\long\def\dh{\hat{D}}%
 
\global\long\def\lage{\mathbf{E}}%
 
\global\long\def\inc#1{\dot{#1}}%
 
\global\long\def\sys{\mathbf{x}}%
 
\global\long\def\curl{\nabla}%
 
\global\long\def\Curl{\nabla_{\refsys}}%
 
\global\long\def\piolaincpush{\boldsymbol{\Sigma}}%
 
\global\long\def\piolaincpushcomp#1{\Sigma_{#1}}%
 
\global\long\def\edlincpush{\check{\mathbf{d}}}%
 
\global\long\def\edlincpushcomp#1{\check{d}_{#1}}%
 
\global\long\def\efincpush{\check{\mathbf{e}}}%
 
\global\long\def\efincpushcomp#1{\check{e}_{#1}}%
 
\global\long\def\elaspush{\boldsymbol{\mathcal{C}}}%
 
\global\long\def\elecpush{\boldsymbol{\mathcal{A}}}%
 
\global\long\def\elaselecpush{\boldsymbol{\mathcal{B}}}%
 
\global\long\def\disgrad{\mathbf{h}}%
 
\global\long\def\disgradcomp#1{h_{#1}}%
 
\global\long\def\trans#1{#1^{\mathrm{T}}}%
 
\global\long\def\phase#1{#1^{\left(p\right)}}%
 
\global\long\def\elecpushcomp#1{\mathcal{A}_{#1}}%
 
\global\long\def\elaselecpushcomp#1{\mathcal{B}_{#1}}%
 
\global\long\def\elaspushcomp#1{\mathcal{C}_{#1}}%
 
\global\long\def\dnh{\aef_{DG}}%
 
\global\long\def\dnhc{\mu\lambda^{2}}%
 
\global\long\def\dnhcc{\frac{\mu}{\lambda^{2}}+\frac{1}{\matper}d_{2}^{2}}%
 
\global\long\def\dnhb{\frac{1}{\matper}d_{2}}%
 
\global\long\def\afreq{\omega}%
 
\global\long\def\dispot{\phi}%
 
\global\long\def\edpot{\varphi}%
 
\global\long\def\kh{\hat{k}}%
 
\global\long\def\afreqh{\hat{\afreq}}%
 
\global\long\def\phasespeed{c}%
 
\global\long\def\bulkspeed{c_{B}}%
 
\global\long\def\speedh{\hat{c}}%
 
\global\long\def\dhth{\dh_{th}}%
 
\global\long\def\bulkspeedl{\bulkspeed_{\lambda}}%
 
\global\long\def\khth{\hat{k}_{th}}%
 
\global\long\def\p#1{#1^{\left(p\right)}}%
 
\global\long\def\maxinccomp#1{\inc{\outer{\sigma}}_{#1}}%
 
\global\long\def\maxcomp#1{\outer{\sigma}_{#1}}%
 
\global\long\def\relper{\matper_{r}}%
 
\global\long\def\sdh{\hat{d}}%
 
\global\long\def\iee{\varphi}%
 
\global\long\def\effectivemu{\tilde{\mu}}%
 
\global\long\def\fb#1{#1^{\left(a\right)}}%
 
\global\long\def\mt#1{#1^{\left(b\right)}}%
 
\global\long\def\phs#1{#1^{\left(p\right)}}%
 
\global\long\def\thc{h}%
 
\global\long\def\state{\mathbf{s}}%
 
\global\long\def\harmonicper{\breve{\matper}}%
 
\global\long\def\kb{k_{B}}%
 
\global\long\def\cb{\bar{c}}%
 
\global\long\def\mb{\bar{\mu}}%
 
\global\long\def\rb{\bar{\rho}}%
 
\global\long\def\wavenumber{k}%
 
\global\long\def\nh{\mathbf{n}}%
 
\global\long\def\mh{\mathbf{m}}%
 
\global\long\def\deflect{\inc x_{2}}%
 
\global\long\def\sdd#1{#1_{2,11}}%
 
\global\long\def\sdddd#1{#1_{2,1111}}%
 
\global\long\def\sd#1{#1_{2,1}}%
 
\global\long\def\sddd#1{#1_{2,111}}%
 
\global\long\def\xdddd#1{#1_{,\xi\xi\xi\xi}}%
 
\global\long\def\xdd#1{#1_{,\xi\xi}}%
 
\global\long\def\xd#1{#1_{,\xi}}%
 
\global\long\def\xddd#1{#1_{,\xi\xi\xi}}%
 
\global\long\def\jm{J_{m}}%
 
\global\long\def\dv{\Delta V}%
 
\global\long\def\ih{\mathbf{i}_{1}}%
 
\global\long\def\kh{\mathbf{i}_{3}}%
 
\global\long\def\jh{\mathbf{i}_{2}}%
 
\global\long\def\etil{E}%
 
\global\long\def\genT{\mathsf{Q}}%
 
\global\long\def\transfer{\mathsf{T}}%
 
\global\long\def\statevec{\mathbf{s}}%
 
\global\long\def\coefvec{\mathbf{c}}%
 
\global\long\def\pressure{p_{0}}%
 
\global\long\def\ncell#1{#1_{\left(n\right)}}%
 
\global\long\def\ydisp{\inc x_{2}}%
 
\global\long\def\ycord{x_{2}}%
 
\global\long\def\pn#1{\ncell{#1}^{\left(p\right)}}%
 
\global\long\def\pnm#1{#1_{\left(n\right)m}^{\left(p\right)}}%
 
\global\long\def\eigen{\alpha}%
 
\global\long\def\xcomp{x_{1}}%
 
\global\long\def\totalT{\mathsf{T_{\mathrm{tot}}}}%
 
\global\long\def\rads{\frac{\mathrm{rad}}{\mathrm{s}}}%
 
\global\long\def\lf{\gamma}%
 
\global\long\def\tf{T_{m}}%
 
\global\long\def\eigenim{\beta}%
 
\global\long\def\bS{\mathsf{S}}%
\global\long\def\dis#1{u_{#1}}%
\global\long\def\refden{\rho_{L}}%
\global\long\def\curden{\rho}%
 
\global\long\def\jump#1{\llbracket#1\rrbracket}%
 
\global\long\def\identity{\mathbf{I}}%
 
\global\long\def\area{\mathrm{d}a}%
 
\global\long\def\areal{\mathrm{d}A}%
 
\global\long\def\refsys{\mathbf{X}}%
 
\global\long\def\Grad{\nabla_{\refsys}}%
 
\global\long\def\grad{\nabla}%
 
\global\long\def\divg{\nabla\cdot}%
 
\global\long\def\Div{\nabla_{\refsys}}%
 
\global\long\def\derivative#1#2{\frac{\partial#1}{\partial#2}}%
 
\global\long\def\aef{\Psi}%
 
\global\long\def\dltendl{\edl\otimes\edl}%
 
\global\long\def\tr#1{\mathrm{tr}#1}%
 
\global\long\def\ii#1{I_{#1}}%
 
\global\long\def\dh{\hat{D}}%
 
\global\long\def\inc#1{\dot{#1}}%
 
\global\long\def\sys{\mathbf{x}}%
 
\global\long\def\curl{\nabla}%
 
\global\long\def\Curl{\nabla_{\refsys}}%
 
\global\long\def\piolaincpush{\boldsymbol{\Sigma}}%
 
\global\long\def\piolaincpushcomp#1{\Sigma_{#1}}%
 
\global\long\def\edlincpush{\check{\mathbf{d}}}%
 
\global\long\def\edlincpushcomp#1{\check{d}_{#1}}%
 
\global\long\def\efincpush{\check{\mathbf{e}}}%
 
\global\long\def\efincpushcomp#1{\check{e}_{#1}}%
 
\global\long\def\elaspush{\boldsymbol{\mathcal{C}}}%
 
\global\long\def\elecpush{\boldsymbol{\mathcal{A}}}%
 
\global\long\def\elaselecpush{\boldsymbol{\mathcal{B}}}%
 
\global\long\def\disgrad{\mathbf{h}}%
 
\global\long\def\disgradcomp#1{h_{#1}}%
 
\global\long\def\trans#1{#1^{\mathrm{T}}}%
 
\global\long\def\phase#1{#1^{\left(p\right)}}%
 
\global\long\def\elecpushcomp#1{\mathcal{A}_{#1}}%
\global\long\def\elaselecpushcomp#1{\mathcal{B}_{#1}}%
\global\long\def\elaspushcomp#1{\mathcal{C}_{#1}}%
 
\global\long\def\dnh{\aef_{DH}}%
\global\long\def\woo{\varsigma}%
\global\long\def\wif{\Lambda}%
\global\long\def\structurefun{S}%
\global\long\def\bondechden{\rho_{b}}%
\global\long\def\surfacebondech{w_{b}}%
\global\long\def\refbondechden{\rho_{B}}%
\global\long\def\refsurfacebondech{w_{B}}%
\global\long\def\incbondechden{\check{\rho}_{b}}%
 
\global\long\def\GmGt{\left(\G-\Gt\right)}%
\global\long\def\woo{\varsigma}%
\global\long\def\fA{\mathbb{\boldsymbol{\mathscr{A}}}}%
\global\long\def\fB{\mathbb{\boldsymbol{\mathscr{B}}}}%
\global\long\def\fC{\mathbb{\boldsymbol{\mathscr{C}}}}%
\global\long\def\hyper{\Psi}%

\global\long\def\piolacomp#1{P_{#1}}%
 
\global\long\def\derivativesec#1#2{\frac{\partial^{2}#1}{\partial#2^{2}}}%

\global\long\def\pressurestrain{M}%
\global\long\def\shearstrain{Q}%
\global\long\def\pressurevel{R}%
\global\long\def\shearvel{S}%

\global\long\def\ipressure{\pressurestrain_{I}}%
\global\long\def\ishear{\shearstrain_{I}}%
\global\long\def\bpressure{\pressurestrain_{B}}%
\global\long\def\bshear{\shearstrain_{B}}%
\global\long\def\ipressurevel{\pressurevel_{I}}%
\global\long\def\ishearvel{\shearvel_{I}}%
\global\long\def\bpressurevel{\pressurevel_{B}}%
\global\long\def\bshearvel{\shearvel_{B}}%

\global\long\def\cpressure{\pressurestrain_{\mathrm{cr}}}%

\global\long\def\upressure{\pressurestrain_{U}}%
\global\long\def\ushear{\shearstrain_{U}}%
\global\long\def\upressurevel{\pressurevel_{U}}%
\global\long\def\ushearvel{\shearvel_{U}}%

\global\long\def\piolaonebyP{\alpha}%
\global\long\def\piolaonebyQ{\beta}%
\global\long\def\piolatwobyP{\gamma}%
\global\long\def\piolatwobyQ{\delta}%
\global\long\def\xbyt{c}%

\global\long\def\shearwavespeed{c_{-}}%
\global\long\def\pressurewavespeed{c_{+}}%

\global\long\def\piolapressure{\sigma}%
\global\long\def\piolashear{\tau}%

\global\long\def\ipiolapressure{\piolapressure_{I}}%
\global\long\def\ipiolashear{\piolashear_{I}}%
\global\long\def\bpiolapressure{\piolapressure_{B}}%
\global\long\def\bpiolashear{\piolashear_{B}}%
\global\long\def\upiolapressure{\piolapressure_{U}}%
\global\long\def\upiolashear{\piolashear_{U}}%

\global\long\def\ii#1{I_{#1}}%
\global\long\def\bulk{\kappa}%
\global\long\def\shear{\mu}%

\global\long\def\speedBS{V_{BS}}%
\global\long\def\speedSU{V_{SU}}%
\global\long\def\speedUP{V_{UP}}%
\global\long\def\speedPI{V_{PI}}%
\global\long\def\speedS{V_{S}}%
\global\long\def\speedP{V_{P}}%

\global\long\def\tr#1{\mathrm{tr}#1}%
 
\global\long\def\det#1{\mathrm{det}#1}%

\title{Smooth waves and shocks of finite amplitude in soft materials}
\author{Ron Ziv and Gal Shmuel\thanks{Corresponding author. Tel.: +1 972 778871613. \emph{E-mail address}:
meshmuel@technion.ac.il (G. Shmuel).}\\
 {\small{}{}{}Faculty of Mechanical Engineering, Technion\textendash Israel
Institute of Technology, Haifa 32000, Israel}\\
 }
\maketitle
\begin{abstract}
Recently developed soft materials exhibit nonlinear wave propagation
with potential applications for energy trapping, shock mitigation
and wave focusing. We address finitely deformed materials subjected
to combined transverse and axial impacts, and study the resultant
nonlinear waves. We determine the dependency of the induced motion
on the impact, pre-deformation and the employed constitutive models.
We analyze the neo-Hookean constitutive model and show it cannot capture
shear shocks and tensile-induced shocks, in contrast with experimental
results on soft materials. We find that the Gent constitutive model
predicts that compressive impact may not be sufficient to induce a
quasi-pressure shock---yet it may induce a quasi-shear shock, where
tensile impact can trigger quasi-pressure shock---and may simultaneously
trigger a quasi-shear shock, in agreement with experimental data.
We show that the tensile impact must be greater than a calculated
threshold value to induce shock, and demonstrate that this threshold
is lowered by application of  pre-shear. 

\end{abstract}

\section{Introduction}

Recent technological advances have promoted the development of soft
materials with reconfigurable properties, capable of undergoing reversible
finite deformations \citep{Bandyopadhyay2015MRS,Truby2016,He2017AFM,KIM2017}.
The nonlinearities associated with these materials give rise to unique
transport properties, that can be exploited for tunable dynamic response,
energy trapping, shock mitigation and wave focusing \citep{kochmann14pre,Shmuel2016JMPS,LINTS2017wm,LUSTIG2018jmps,Giammarinaro2018PRapplied}.
These potential applications have led to a revived scientific interest
in nonlinear wave propagation of elastic continua \citep{raney16,Xin2016sr,deng2017}.
The pioneering theoretical work in the field is mainly attributed
to Carroll \citeyearpar{carroll1967some,Carroll1974,Carroll1978},
followed by the works of \citet{BOULANGER1992}, \citet{RAJAGOPAL1998ijnlm}
and more recently \citet{destrade2005finite}. The focus of these
studies is on the existence of harmonic waves with finite amplitude
and constant waveforms. Such waves are the exception rather than the
rule when nonlinearities are accounted for. Our focus is on finite
amplitude smooth waves whose waveform changes in soft materials under
impact, which distinguishes what follows from the foregoing studies.
Specifically, we study their coalescence to \emph{shocks}---propagating
surfaces of discontinuity in the governing fields. 

An excellent cover of the research on shocks in solids is given by
\citet{davison2008book}. One of the central works on shocks in soft
materials was by \citet{knowles2002impact}, who analyzed a one-dimensional
bar with cubic stress-strain relation under tensile impact. Using
the concept of \emph{thermodynamic driving force}, Knowles theoretically
showed that tensile shocks emerge when the impact is sufficiently
strong. \citet{NIEMCZURA2011423,NIEMCZURA2011442} have designed and
executed corresponding experiments using strips of latex and nitrile
rubber. Our objectives are to (\emph{i}) also account for transverse
displacements and their coupling with the axial displacements; (\emph{ii})
comprehensively study the effect that combined pre-shear and pre-stretch
have on finite amplitude waves induced by combined shear and axial
impacts; (\emph{iii}) characterize the dependency of the resultant
waves on the constitutive models. 

Finite amplitude shocks in similar settings were addressed by several
researchers, whose objectives are different that the objectives in
this work. \citet{DAVISON1966249} proposed a theory with admissibility
conditions for shocks to obtain general formulas and demonstrated
his theory using an example problem. \citet{ABOUDI1973ijss} developed
a finite difference-based scheme to solve the equations governing
impact-induced nonlinear waves. \citet{li1982ijss} introduced the
concept of \emph{stress paths} to determine general solutions, and
exemplified their approach using second order isotropic materials.
More recently, \citet{Scheidler2000aip} derived \emph{universal}
relations\textemdash independent of the specific constitutive relation\textemdash between
the governing fields and the wave velocities. 

In the sequel, we employ the theory of \citet{DAVISON1966249} to
obtain and analyze explicit solutions for plane waves of finite amplitude
in semi-infinite soft materials under different pre-deformations and
impacts. We employ the two most prominent constitutive models for
soft materials---the compressible neo-Hookean and Gent models---to
describe the stress-strain relation \citep{gent96rc&t,PUGLISI201517}.
While both models account for finite deformations, the Gent model
incorporates an additional nonlinearity, aimed at capturing the stiffening
that rubber exhibits due to the limited extensibility of its polymer
chains. Interestingly, the Gent model is also useful to describe soft
tissues, which stiffen due to the stretching of their collagen fibers
\citep{Horgan201ijnlm}. This nonlinearity, in turn, significantly
affects the response of the material to impact and formation of shocks,
as we show in what follows. 

In the limit of small strains, the axial and transverse impact of
a semi-infinite medium excites two types of waves, namely, shear and
pressure waves. These waves do not interact with each other, and their
propagation is independent of the specific loading program and initial
state of the material; shocks propagate with the same velocities.
By contrast, these waves are generally coupled in finite elasticity,
owing to the Poynting effect \citep{Cioroianu2013PRE,Horgan2017prsa};
their velocity depends on the constitutive nonlinearity, initial deformation
and loading program. In this work, we thoroughly study the effect
of these parameters on the propagation of finite amplitude waves. 

The findings from the neo-Hookean model are relatively similar to
the linear theory, as in the considered settings the model does not
capture the coupling between axial and transverse displacements. Specifically,
smooth shear waves propagate at a constant velocity as in the linear
theory, and cannot evolve to shock. Smooth pressure waves, however,
propagate in a velocity that varies as a function of the axial displacement
gradient, and coalesce into shock only when the axial impact tends
to compress the material, \emph{e.g.}, when a pre-stretched material
is released. Accordingly, there are no tensile-induced shocks in neo-Hookean
materials. The neo-Hookean predictions are therefore incompatible
with experimental data on shear shocks and tensile-induced shocks
in soft materials \citep{Catheline2003PRL,NIEMCZURA2011442}. 

The predictions of the Gent model are significantly different and
more interesting than the neo-Hookean model. Firstly, the model captures
the coupling between the transverse and axial motions. We refer to
waves at which the only displacement that retained upon linearization
is the transverse (resp.~axial) displacement as \emph{quasi}-shear
(resp.~\emph{quasi}-pressure). Their velocities depend both on the
initial state and impact program. We characterize this dependency,
and determine when they coalesce into shocks. Interestingly, we find
that compressive impact may not result in a quasi-pressure shock---yet
it may excite a quasi-shear shock, while tensile impact can induce
a quasi-pressure shock and a quasi-shear shock at the same time. We
show that the tensile impact must be greater than a threshold value
to induce shock, and demonstrate that this threshold is lowered when
applying pre-shear. Contrary to the neo-Hookean model, the Gent model
is able to recover the aforementioned experimental results. 

The study is presented in the following order. Sec.$\ $\ref{sec:Problem-statement}
contains the mathematical description of the problem, and its general
resolution for smooth and shock wave solutions \citep{DAVISON1966249}.
The generic solution is specialized in Secs.~\ref{sec:neo-Hookean}
and \ref{sec:Gent}, respectively, to the compressible neo-Hookean
and Gent models, where we characterize the dependency of the wave
velocity on the loading conditions and model parameters, and qualitatively
analyze the criterion for shock. Sec.~\ref{sec:atRest} specializes
and quantifies this criterion in terms of the loading parameters,
when the material is initially unstrained. Sec.~\ref{sec:strained}
extends the study of this criterion to finitely strained materials.
We conclude this paper with a summary of our results and comments
on future work in Sec.~\ref{Conclusion}.

\section{Problem statement and method of solution \label{sec:Problem-statement}}

\begin{figure}[t]
\centering\includegraphics[width=1\textwidth]{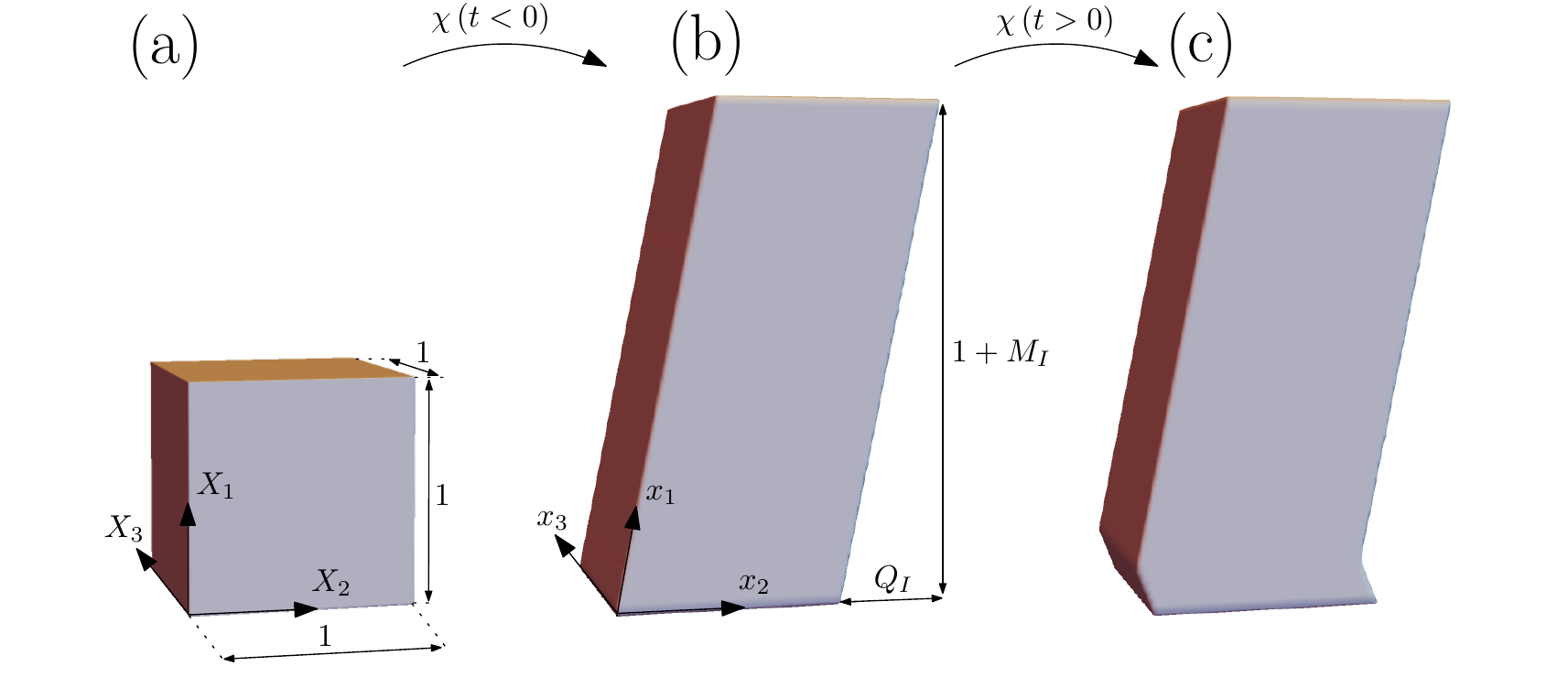}\protect\caption{A unit cube within a semi-infinite body in the (a) reference configuration;
(b) pre-deformed configuration under a uniform shear $\protect\ishear$
and axial displacement gradient $\protect\ipressure$. (c) Illustrative
impact-induced surface of discontinuity at $t>0$.}

{\small{}{}{}\label{body}} 
\end{figure}

The general treatment of the problem using a semi-inverse approach
dates back to \citet{DAVISON1966249}, which we revisit here for completeness.
Consider a semi-infinite soft and compressible material occupying
the region $X_{1}\geqslant0$ in a reference configuration. The material
is hyperelastic, such that the $1^{\mathrm{st}}$ Piola-Kirchhoff
stress $\piola$ is derived from a strain energy function $\hyper$.
Let $\map$ denote the deformation of material points from a reference
coordinate $\mathbf{X}$ to the current coordinate $\mathbf{x}$,
where $\defgrad=\grad_{\mathbf{X}}\deformation$ is the deformation
gradient, then $\piola=\derivative{\hyper}{\dg}$. We focus on (initially)
isotropic materials, for which
\begin{equation}
\piola=\alpha_{1}\defgrad+\alpha_{2}\defgrad\defgradT\defgrad+\alpha_{3}\defgradTi\label{eq:piolastressstrain}
\end{equation}
for some response functions $\alpha_{i}$ that depend on $\hyper$.
At the initial state, the material is sheared and strained along $X_{1}$
homogeneously. Subsequently, the body is subjected to a combination
of transverse and axial impact at its boundary. Accordingly, the continuous
mapping $\map$ (Fig. \ref{body})
\begin{equation}
x_{1}=X_{1}+\dis 1\left(X_{1},t\right),\;x_{2}=X_{2}+\dis 2\left(X_{1},t\right),\;x_{3}=X_{3},\label{eq:deformation}
\end{equation}
is subjected to the initial and boundary conditions

\begin{equation}
\begin{aligned} & \derivative{\dis 1}{X_{1}}\left(0,t\right)=\begin{cases}
\ipressure & t\leq0\\
\bpressure & t>0
\end{cases}\eqqcolon M\left(0,t\right),\;\derivative{\dis 2}{X_{1}}=\begin{cases}
\ishear & t\leq0\\
\bshear & t>0
\end{cases}\eqqcolon\shearstrain\left(0,t\right),\\
 & \derivative{\dis 1}t\left(0,t\right)=\begin{cases}
0 & t\leq0\\
\bpressurevel & t>0
\end{cases}\eqqcolon\pressurevel\left(0,t\right),\;\derivative{\dis 2}t=\begin{cases}
0 & t\leq0\\
\bshearvel & t>0
\end{cases}\eqqcolon\shearvel\left(0,t\right).
\end{aligned}
\label{eq:boundaryInitialstrain}
\end{equation}
where we recall that at $t=0$ the fields are homogeneous. The outstanding
problem is to determine the resultant motion $u_{i}\left(X_{1},t\right)$
for $t>0$.

\subsection{Smooth and acceleration wave solutions}

Assuming $\map$ is continuously differentiable twice almost everywhere,
the differential equations governing the problem (except at singular
surfaces) are
\begin{equation}
\Refdiv{\mathbf{\piola}}=\refden\deformation_{,tt},\label{eq:eom}
\end{equation}
where $\refden=J\curden$, $J=\det{\defgrad}$ and $\curden$ is the
current mass density. Eq.$\ $\eqref{eq:eom} takes the form 

\begin{equation}
\begin{aligned}\piolacomp{11,1}=\refden\derivativesec{x_{1}}t,\;\piolacomp{21,1}=\refden\derivativesec{x_{2}}t,\end{aligned}
\label{eq:motion}
\end{equation}
for the motion considered in Eq.$\ $\eqref{eq:deformation}, and
the constitutive form \eqref{eq:piolastressstrain}. We use the variables
$\pressurevel,\pressurestrain,\shearstrain$ and $\shearvel$ defined
in Eq.$\ $\eqref{eq:boundaryInitialstrain} to reduce the order of
Eq.$\ $\eqref{eq:motion} and obtain

\begin{equation}
\begin{aligned} & \piolacomp{11,1}=\refden\pressurevel{}_{,t},\;\piolacomp{21,1}=\refden\shearvel_{,t},\\
 & \pressurestrain_{,t}=\pressurevel_{,1},\;\shearstrain_{,t}=\shearvel_{,1}.
\end{aligned}
\label{eq:orderreduce}
\end{equation}
By application of the chain rule, we rewrite Eqs.$\ $\eqref{eq:orderreduce}$_{1,2}$
as

\begin{equation}
\begin{aligned} & \piolaonebyP\pressurestrain{}_{,1}+\piolaonebyQ\shearstrain_{,1}=\pressurevel_{,t},\\
 & \piolatwobyP\pressurestrain_{,1}+\piolatwobyQ\shearstrain_{,1}=\shearvel_{,t},
\end{aligned}
\label{eq:simplifiedEq}
\end{equation}
where

\begin{equation}
\piolaonebyP=\frac{1}{\refden}\derivative{\piolacomp{11}}{\pressurestrain},\;\piolaonebyQ=\frac{1}{\refden}\derivative{\piolacomp{11}}{\shearstrain},\;\piolatwobyP=\frac{1}{\refden}\derivative{\piolacomp{21}}{\pressurestrain},\;\piolatwobyQ=\frac{1}{\refden}\derivative{\piolacomp{21}}{\shearstrain}.\label{eq:defineCoeff}
\end{equation}
We firstly seek smooth wave solutions that depend on a single independent
variable. We set this variable to be $\xbyt=X_{1}/t$, and obtain

\begin{equation}
\begin{aligned} & \piolaonebyP\pressurestrain_{,\xbyt}+\piolaonebyQ\shearstrain_{,\xbyt}+\xbyt\pressurevel_{,\xbyt}=0,\;\piolatwobyP\pressurestrain_{,\xbyt}+\piolatwobyQ\shearstrain_{,\xbyt}+\xbyt\shearvel_{,\xbyt}=0,\\
 & \pressurevel_{,\xbyt}+Z\pressurestrain_{,\xbyt}=0,\;\shearvel_{,\xbyt}+Z\shearstrain_{,\xbyt}=0,
\end{aligned}
\label{eq:EqOneVariable}
\end{equation}
from Eqs.$\ $\eqref{eq:orderreduce}$_{3,4}$ and \eqref{eq:simplifiedEq}.
Substitution of Eqs.$\ $\eqref{eq:EqOneVariable}$_{3}$ and \eqref{eq:EqOneVariable}$_{4}$
into Eqs.$\ $\eqref{eq:EqOneVariable}$_{1}$ and \eqref{eq:EqOneVariable}$_{2}$,
respectively, provides

\begin{equation}
\begin{aligned}\left(\piolaonebyP-\xbyt^{2}\right)\pressurestrain_{,\xbyt}+\piolaonebyQ\shearstrain_{,\xbyt}=0,\\
\piolatwobyP\pressurestrain_{,\xbyt}+\left(\piolatwobyQ-\xbyt^{2}\right)\shearstrain_{,\xbyt}=0.
\end{aligned}
\label{eq:PQequations}
\end{equation}
These equations have a non-trivial solution only if

\begin{equation}
\xbyt^{4}-\left(\piolaonebyP-\piolatwobyQ\right)^{2}\xbyt^{2}+\piolaonebyP\piolatwobyQ-\piolaonebyQ\piolatwobyP=0.\label{eq:characteristicEquation}
\end{equation}
\floatsetup[figure]{style=plain,subcapbesideposition=top}
\begin{figure}[!t]
\begin{raggedright}
\sidesubfloat[]{\includegraphics[scale=0.65]{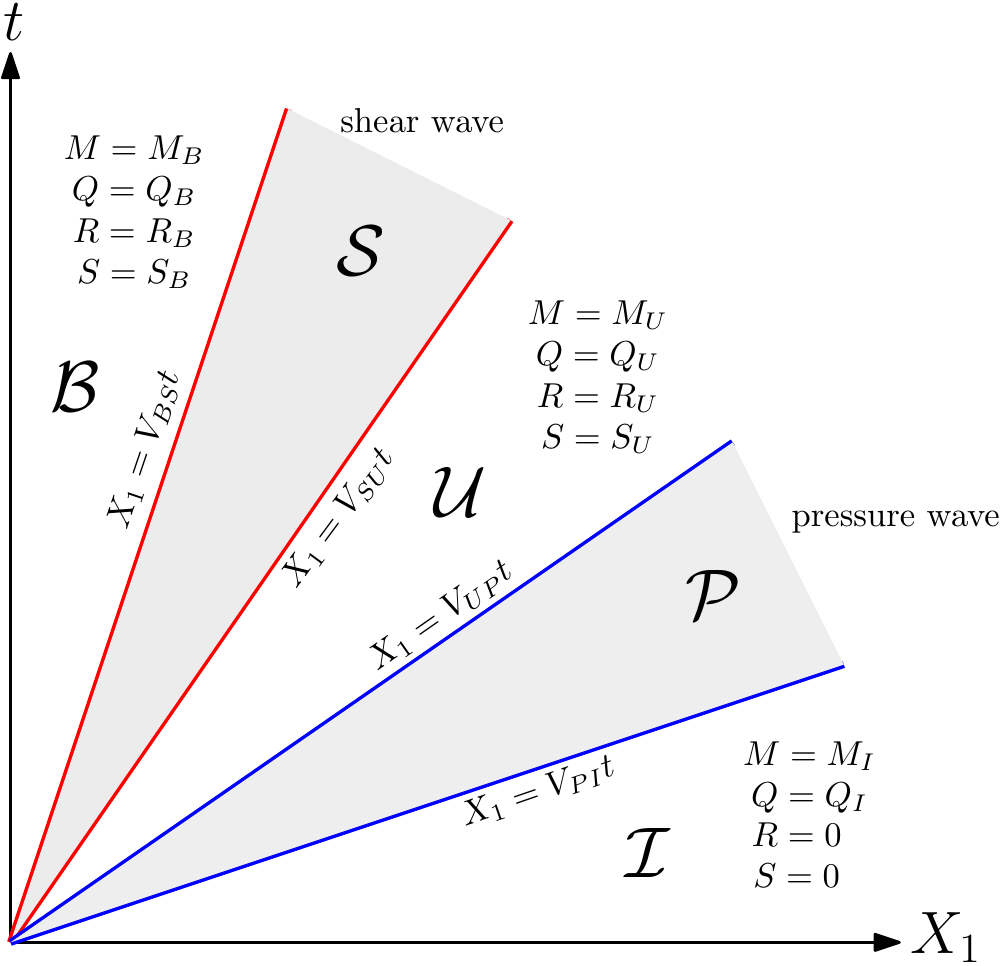}}  $\;\;$\sidesubfloat[]{
\includegraphics[scale=0.65]{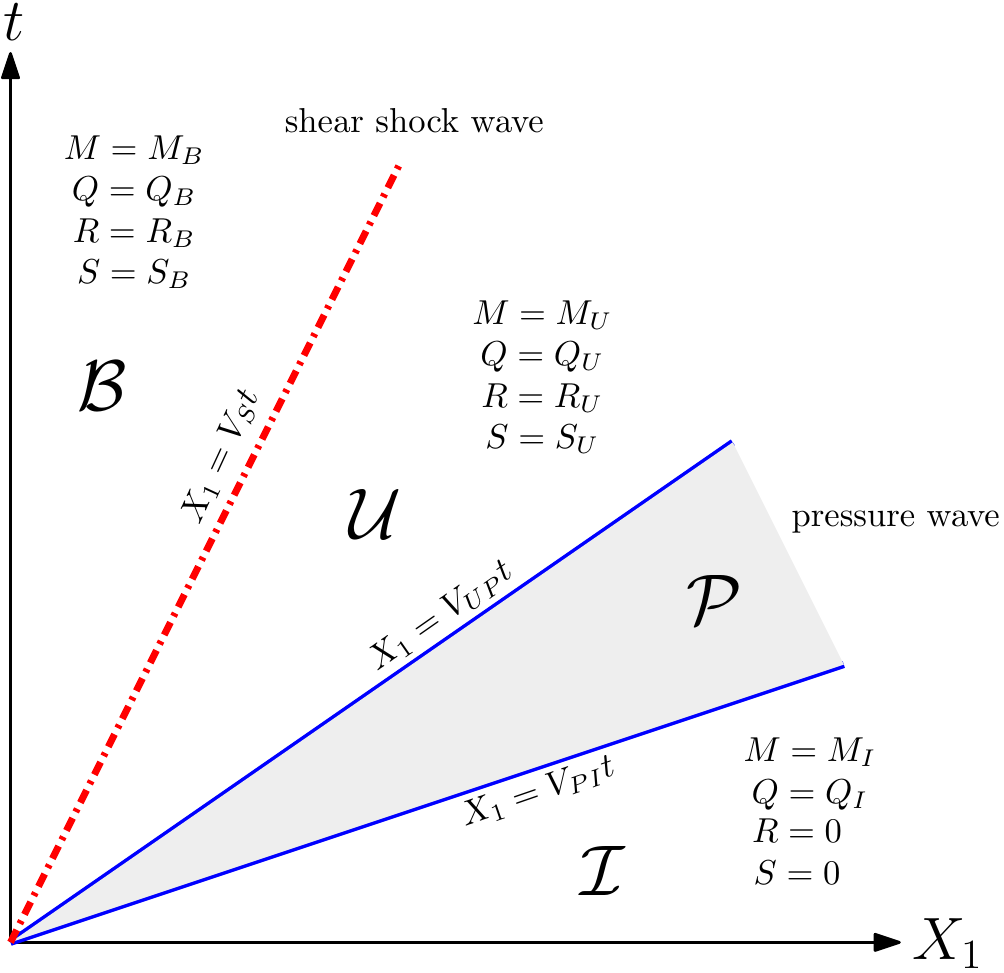}}  
\par\end{raggedright}
\raggedright{}\sidesubfloat[]{\includegraphics[scale=0.65]{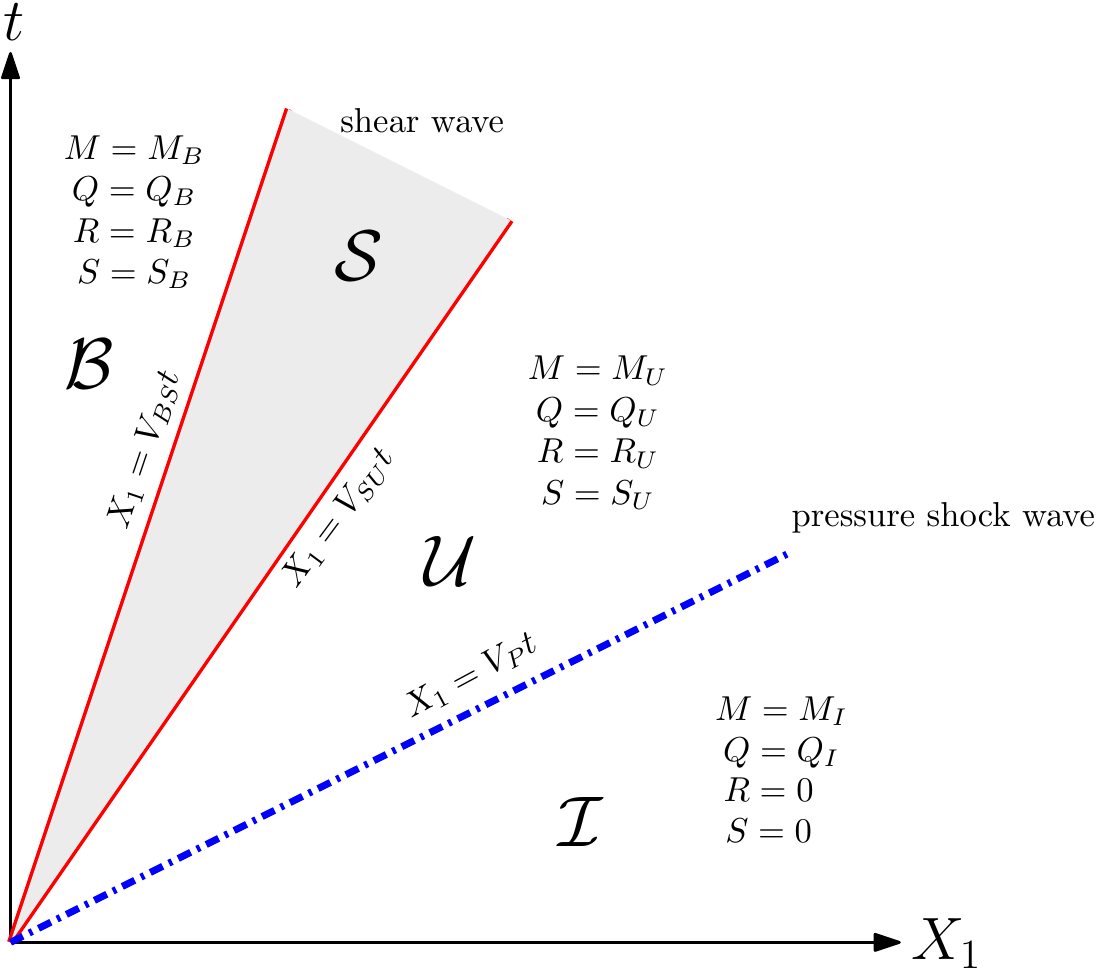}}$\;\;$\sidesubfloat[]{\includegraphics[scale=0.65]{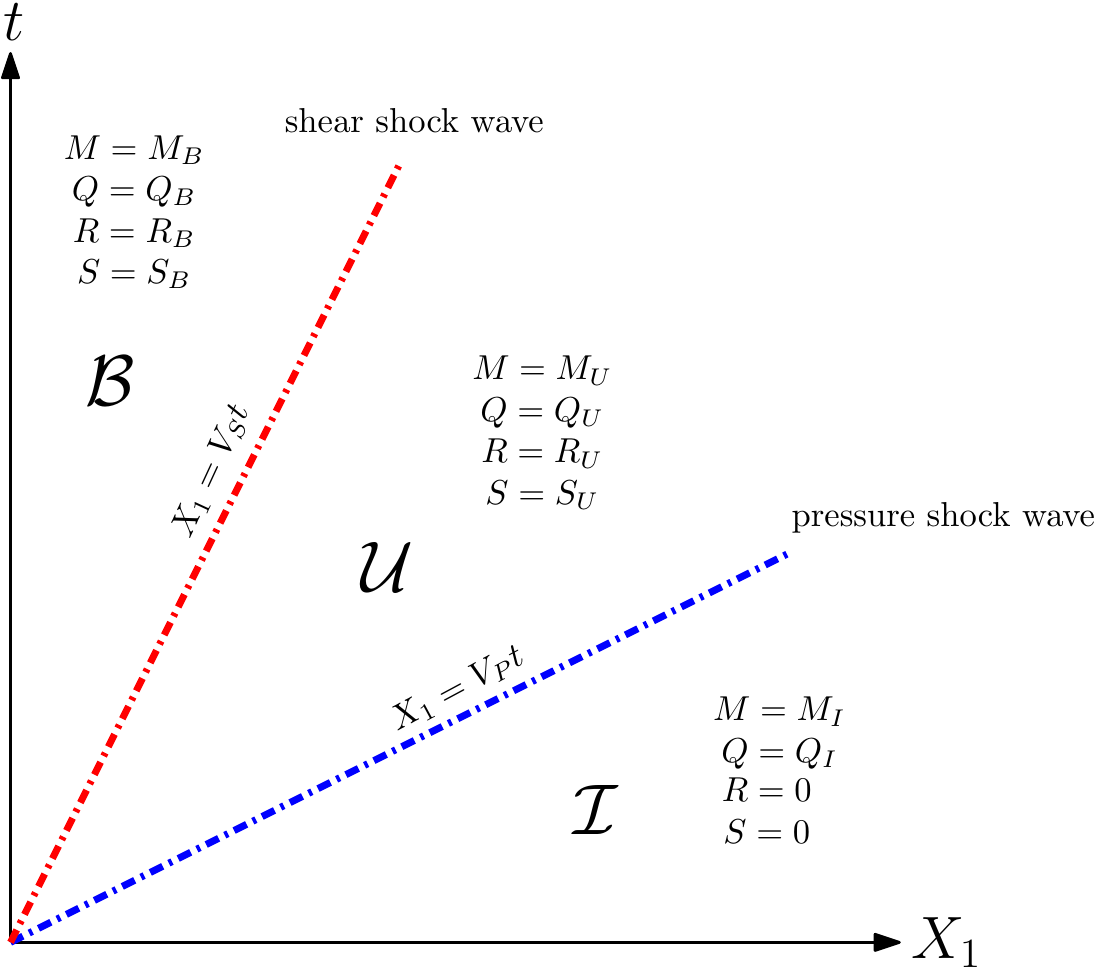}}\caption{Characteristic curves as solutions to boundary and initial conditions.
For monotonic $c$, the solution to the quasi-shear and quasi-pressure
waves is one of the following combinations: (a) smooth-smooth, (b)
shock-smooth, (c) smooth-shock, (d) shock-shock, depending on the
boundary and initial conditions. Dot-dashed and continuous curves
correspond to shock and acceleration waves solutions, respectively.
Smooth waves are denoted in gray.\label{fig:simpleAndshockWaves}}
\end{figure}
Solving for $\xbyt$ gives the characteristic wave velocities\footnote{These velocities are \emph{Lagrangian}, measured relative to the reference
coordinate $X_{1}$. The \emph{Eulerian} velocities, measured with
respect to the moving coordinate $x_{1}$, are $\left(1+\pressurestrain\right)c_{\pm}$
\citep{davison2008book}.}

\begin{equation}
\xbyt^{2}=\frac{1}{2}\left[\piolaonebyP+\piolatwobyQ\pm\sqrt{\left(\piolaonebyP-\piolatwobyQ\right)^{2}+4\piolaonebyQ\piolatwobyP}\right]\eqqcolon c_{\pm}^{2}.\label{eq:wavespeeds}
\end{equation}
In the limit of linear elasticity, the velocities $\pressurewavespeed$
and $\shearwavespeed$ reduce to the velocities of pressure and shear
waves, respectively. In this limit $\pressurewavespeed=\piolaonebyP,\ \shearwavespeed=\piolatwobyQ,\ \beta\gamma=0$,
and accordingly there is no coupling through the equations of motion
between the corresponding waves. 

We refer to the slow and fast smooth waves associated with $\shearwavespeed$
and $\pressurewavespeed$ as \emph{quasi}-shear and \emph{quasi}-pressure
waves, respectively. We assume that these waves spatially expand in
the course of propagation, as illustrated in Fig.$\ $\ref{fig:simpleAndshockWaves}(a)
by the gray regions. We further assume that fields in the white cones
separating these waves and the rays $t=0$ and $X_{1}=0$ are uniform.
The cone that is bounded by $t=0$ (resp.$\ $$X_{1}=0$) is denoted
by $\mathcal{B}$ (resp.$\ $$\mathcal{I}$). The middle cone is denoted
by $\mathcal{U}$. The fields in these regions are denoted with the
subscripts $B,I$ and $U$. To determine the velocity field in between
the front ($X_{1}=\speedSU t$) and back ($X_{1}=\speedBS t$) characteristics
of the quasi-shear wave, we substitute $\shearwavespeed$ back into
Eq.$\ $\eqref{eq:PQequations} and obtain

\begin{equation}
\frac{\mathrm{d}\pressurestrain}{\mathrm{d}\shearstrain}=-\frac{\piolaonebyQ}{\piolaonebyP-\shearwavespeed^{2}}.\label{eq:shearRegionEq}
\end{equation}
These characteristics define surfaces of discontinuity for the second
derivatives of $u_{i}$, and are termed \emph{acceleration} waves.
The first derivatives are continuous, and thus Eq.$\ $\eqref{eq:shearRegionEq}
is subjected to a \emph{compatibility} condition in the form of the
continuity of $\pressurestrain$, such that $\pressurestrain\left(\shearstrain=\bshear\right)=\bpressure$,
where we recall that $\shearwavespeed$ is a function of $\pressurestrain$
and $\shearstrain$. Similarly, to determine the velocity field in
between the front ($X_{1}=\speedPI t$) and back ($X_{1}=\speedUP t$)
characteristics of the quasi-pressure wave we substitute $\pressurewavespeed$
back into Eq.$\ $\eqref{eq:PQequations} and obtain
\begin{equation}
\frac{\mathrm{d}\shearstrain}{\mathrm{d}\pressurestrain}=-\frac{\piolaonebyP-\pressurewavespeed^{2}}{\piolaonebyQ},\label{eq:pressureRegionEq}
\end{equation}
subjected to the compatibility condition $\shearstrain\left(\pressurestrain=\ipressure\right)=\ishear$
for the acceleration waves. Let $\shearstrain_{+}$ (resp.$\ $$\shearstrain_{-}$)
denote the value of $\shearstrain$ at the back (resp.$\ $front)
of the quasi-pressure (resp.$\ $shear) wave; the continuity condition
between the shear and pressure waves is then reads $\shearstrain_{+}=\shearstrain_{-}\eqqcolon\ushear$.

\subsection{Shock wave solutions}

Smooth waves bounded by acceleration waves evolve only when the calculated
velocity at the back of the wave is smaller than the velocity at the
front of the wave, and changes monotonically in between. This requirement
can be written as
\begin{equation}
\begin{aligned} & \speedBS<\shearwavespeed\left(\shearstrain\right)<\speedSU,\;\frac{\mathrm{d}\shearwavespeed}{\mathrm{d}\shearstrain}\neq0\;\;\forall\shearstrain\in\left(\bshear,\ushear\right),\\
 & \speedUP<\pressurewavespeed\left(\pressurestrain\right)<\speedPI,\;\frac{\mathrm{d}\pressurewavespeed}{\mathrm{d}\pressurestrain}\neq0\;\;\forall\pressurestrain\in\left(\upressure,\ipressure\right).
\end{aligned}
\label{eq:simplewaverequirement}
\end{equation}
Shocks evolve when conditions \eqref{eq:simplewaverequirement} fail
to hold, and define a surface of discontinuity for all the governing
fields except the displacements. We restrict attention to the cases
when either $\speedBS<\shearwavespeed\left(\shearstrain\right)<\speedSU$
and/or $\speedBS<\shearwavespeed\left(\shearstrain\right)<\speedSU$
are violated. The integral form of Eq.$\ $\eqref{eq:eom} is then
used to derive the jump conditions

\begin{equation}
\begin{aligned}\refden\left\llbracket \pressurevel\right\rrbracket V+\left\llbracket \piolacomp{11}\right\rrbracket =0,\; & \left\llbracket \pressurestrain\right\rrbracket V+\left\llbracket \pressurevel\right\rrbracket =0,\\
\refden\left\llbracket \shearvel\right\rrbracket V+\left\llbracket \piolacomp{21}\right\rrbracket =0,\; & \left\llbracket \shearstrain\right\rrbracket V+\left\llbracket \shearvel\right\rrbracket =0;
\end{aligned}
\label{eq:jumpCond}
\end{equation}
here, $V$ is the shock velocity and $\left\llbracket \circ\right\rrbracket $
is the jump in $\left(\circ\right)$ between the regions ahead and
behind the shock. Specialization of Eq.$\ $\eqref{eq:jumpCond} to
the loading \eqref{eq:boundaryInitialstrain} delivers the following
conditions for the quasi-shear shock and quasi-pressure shock waves

\begin{equation}
\begin{aligned} & \left\llbracket P_{11}\right\rrbracket _{UB}\left\llbracket \shearstrain\right\rrbracket _{UB}-\left\llbracket P_{21}\right\rrbracket _{UB}\left\llbracket \pressurestrain\right\rrbracket _{UB}=0,\\
 & \left\llbracket P_{11}\right\rrbracket _{IU}\left\llbracket \shearstrain\right\rrbracket _{IU}-\left\llbracket P_{21}\right\rrbracket _{IU}\left\llbracket \pressurestrain\right\rrbracket _{IU}=0,
\end{aligned}
\label{eq:reducedJumpCond}
\end{equation}
respectively, where paired subscripts of $\left\llbracket \circ\right\rrbracket $
denote that the jump is between the corresponding regions; each of
these equations provides a connection between $\ushear$ and $\upressure$.
The shock velocities resulting from Eq.$\ $\eqref{eq:jumpCond} are

\begin{equation}
\speedS=\sqrt{\frac{\left\llbracket P_{21}\right\rrbracket _{UB}}{\refden\left\llbracket \shearstrain\right\rrbracket _{UB}}},\;\speedP=\sqrt{\frac{\left\llbracket P_{11}\right\rrbracket _{IU}}{\refden\left\llbracket \pressurestrain\right\rrbracket _{IU}}},\label{eq:pressureshockspeed}
\end{equation}
where $\speedS$ and $\speedP$ correspond to the quasi-shear and
quasi-pressure shock velocities, respectively. Depending on the loading
program and the specific constitutive behavior, four combinations
of shear and pressure waves can develop, \emph{i.e.}, smooth-smooth
(Fig.$\ $\ref{fig:simpleAndshockWaves}a), shock-smooth (Fig.$\ $\ref{fig:simpleAndshockWaves}b),
smooth-shock (Fig.$\ $\ref{fig:simpleAndshockWaves}c), and shock-shock
(Fig.$\ $\ref{fig:simpleAndshockWaves}d). In general, the theory
does not identify \emph{a priori} which combination takes place. The
course taken is to use a semi-inverse method of solution, \emph{i.e.},
firstly assume smooth-smooth solutions, and examine the compatibility
of the corresponding set of equations. If the compatibility fails
to hold, we assume that shock-smooth waves propagate, and examine
the compatibility of the corresponding equations. This process is
continued until we find a compatible set and identify the combination
that takes place. The foregoing combinations cover all the possibilities
when $c$ is monotonic. 

\section{Analysis of compressible neo-Hookean materials\label{sec:neo-Hookean}}

The study of smooth waves and shocks of finite amplitude in neo-Hookean
and Gent materials is carried out next, using the theory in Sec.~\ref{sec:Problem-statement}.
We begin with the neo-Hookean model

\begin{equation}
\hyper=\frac{\mu}{2}\left(\ii 1-3\right)-\mu\ln J+\left(\frac{\bulk}{2}-\frac{\mu}{3}\right)\left(J-1\right)^{2},
\end{equation}
where $\ii 1=\tr{\defgradT\defgrad}$, and $\bulk$ and $\mu$ correspond
to the bulk and shear moduli, respectively, in the limit of small
strains. For this model, the velocities obtained from Eq.$\ $\eqref{eq:wavespeeds}
are

\begin{equation}
\shearwavespeed=\piolatwobyQ=\sqrt{\frac{\mu}{\refden}},\;\pressurewavespeed=\piolaonebyP=\sqrt{\frac{\bulk}{\refden}+\frac{\mu\left(\pressurestrain^{2}+2\pressurestrain+4\right)}{3\refden\left(\pressurestrain+1\right)^{2}}}.\label{eq:wavespeedNeo}
\end{equation}
Notably, since $\piolaonebyQ\piolatwobyP=0$ the velocities and corresponding
waves are not coupled. Stated differently, the axial displacement
gradient $\pressurestrain$ and shear field $\shearstrain$ evolve
independently in the material.

\floatsetup[figure]{style=plain,subcapbesideposition=top}
\begin{figure}[!t]
\begin{raggedright}
\centering\sidesubfloat[]{\includegraphics[scale=0.35]{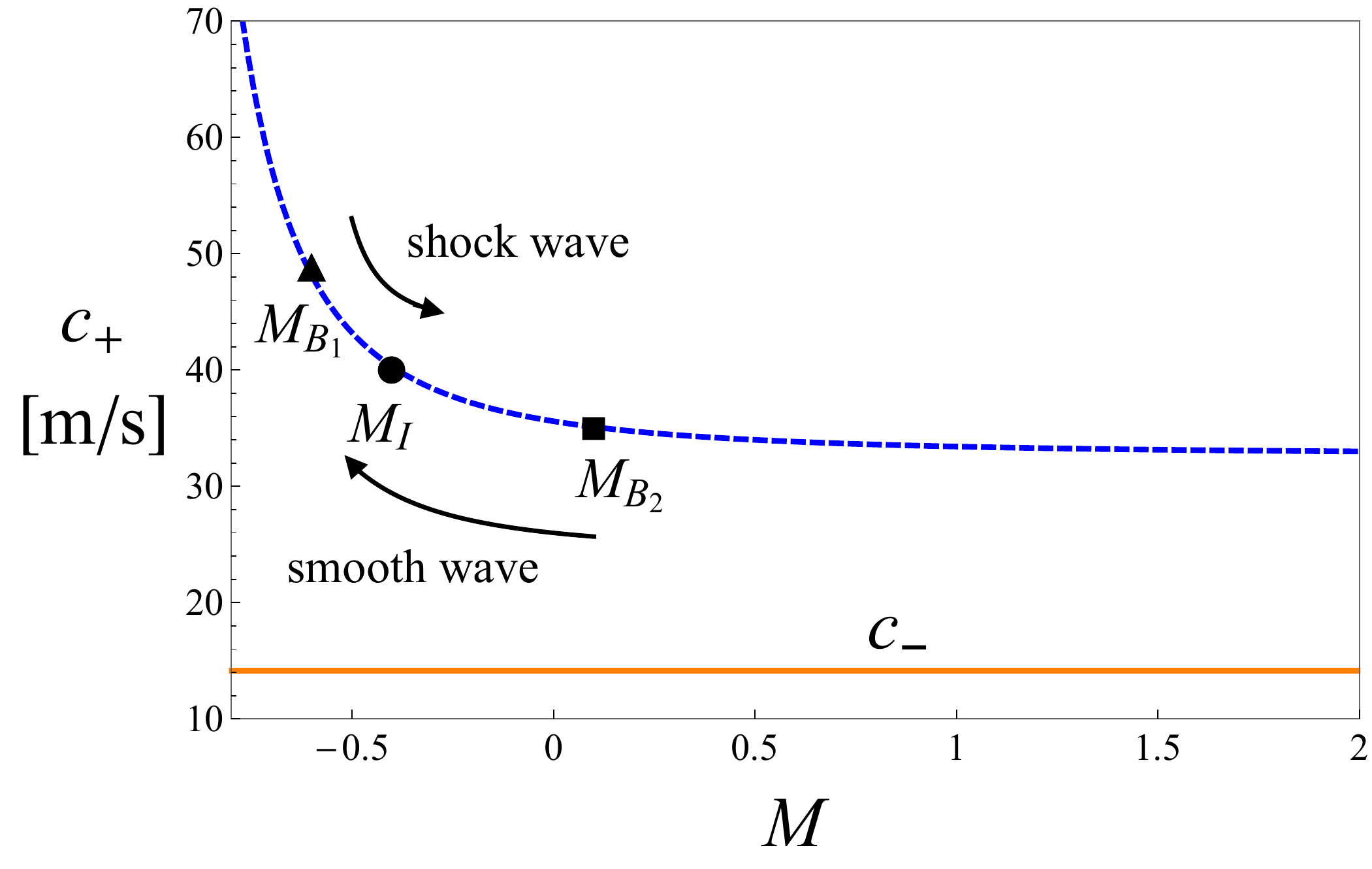}}  $\;\;\;$\sidesubfloat[]{\includegraphics[scale=0.35]{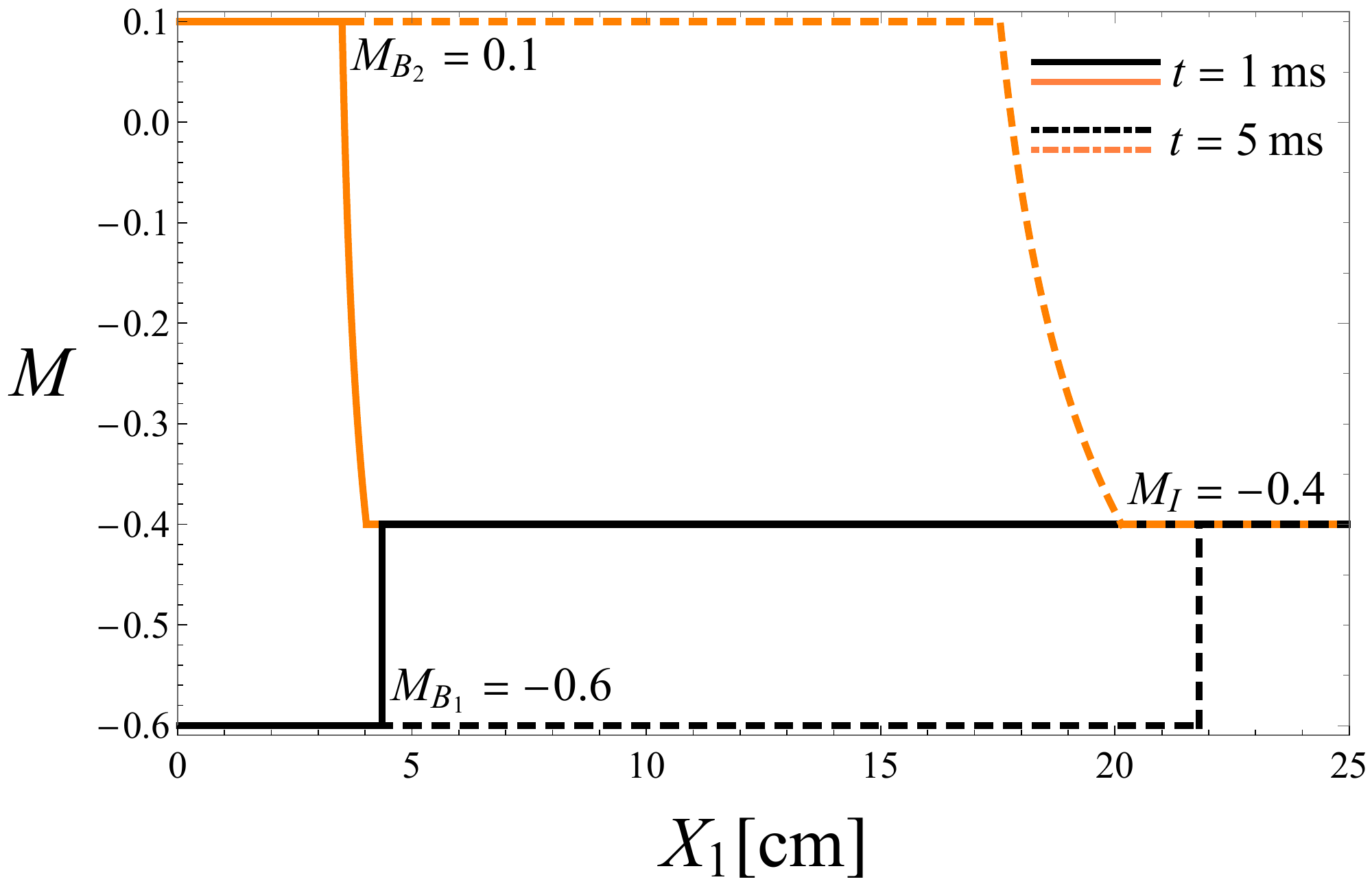}}
\par\end{raggedright}
\caption{\label{fig:a4Neo}(a) Pressure (dashed-blue curve) and shear (orange
line) wave velocities as functions of $\protect\pressurestrain$,
for a neo-Hookean material with the material parameters \eqref{eq:parameters}.
The triangle ($M_{B_{1}}$) and square ($M_{B_{2}}$) marks correspond
to boundary conditions that result in smooth and shock waves, respectively,
when the initial condition is $\protect\ipressure=-0.4$ (circle mark).
(b) Corresponding distributions along $X_{1}$ of the axial displacement
gradient $\protect\pressurestrain$ for $M_{B_{1}}$ (black) and $M_{B_{2}}$
(orange) at $t=1\mathrm{ms}$ (solid curves) and $t=5\mathrm{ms}$
(dashed curves).}
\end{figure}
The neo-Hookean model predicts that (pure) smooth shear waves propagate
with a constant velocity, and hence cannot evolve to shock. Therefore,
the model is incompatible with experimental demonstrations  of shear
shocks in soft solids \citep{Catheline2003PRL,Jacob2007JASA}.

The neo-Hookean model predicts that (pure) pressure waves propagate
as smooth waves with the velocity $\pressurewavespeed$ only if $\bpressure>\ipressure$,
since $\pressurewavespeed$ is a monotonically decreasing function
of $\pressurestrain$. Mechanically speaking, the mathematical condition
implies that only compressive impact leads to shock. Accordingly,
the neo-Hookean model is incapable of recovering experimental data
of tensile-induced shocks in soft materials \citep{kolsky1969production,NIEMCZURA2011442}.
The dependency of $\pressurewavespeed$ on $\pressurestrain$ is exemplified
in Fig.$\ $\ref{fig:a4Neo}(a), which shows $\pressurewavespeed$
as a function of $\pressurestrain$ (dashed-blue curve), for the representative
parameters 
\begin{equation}
\bulk=1\mathrm{MPa},\;\shear=200\mathrm{kPa},\;\refden=1000\mathrm{\mathrm{kg/m^{3}}};\label{eq:parameters}
\end{equation}
for comparison, the shear wave velocity is also displayed (orange
line). To complete the solution, we need to determine $\upressure$
and $\ushear$; owing to the decoupling of the velocities, we have
that $\upressure=\bpressure$ and $\ushear=\ishear$. If $\bpressure<\ipressure$,
then the pressure wave will coalesce into shock with the velocity
\eqref{eq:pressureshockspeed}$_{2}$.

By way of example, we examine a material at the initial compression
state $\ipressure=-0.4$ and consider two different impacts at the
boundary, namely $M_{B_{1}}=-0.6$ and $M_{B_{2}}=0.1$. These impacts
are indicated in panel \ref{fig:a4Neo}(a) by the triangle $\left(M_{B_{1}}\right)$
and square $\left(M_{B_{2}}\right)$ marks. The former propagates
as shock with the velocity $\speedP=43.58\mathrm{m/s}$, since $M_{B_{1}}<\ipressure$.
The latter propagates as a smooth wave with the velocity range $35.01\mathrm{m/s}<\pressurewavespeed\left(\pressurestrain\right)<40.27\mathrm{m/s}$.
The corresponding distributions along $X_{1}$ of $\pressurestrain$
for the two boundary conditions at $t=1\mathrm{ms}$ (solid curves)
and $t=5\mathrm{ms}$ (dashed curves) are shown in Fig.$\ $\ref{fig:a4Neo}(b).
We observe that indeed $M_{B_{1}}$(black) is associated with the
propagation of strain discontinuity. Conversely, applying $M_{B_{2}}$
(orange) yields smooth spreading of strain bounded between acceleration
waves, identified by discontinuities in the derivative of $\pressurestrain$.

\section{Analysis of compressible Gent materials\label{sec:Gent}}

The second model we employ is the compressible Gent model \citep{gent96rc&t}.
This model was developed to capture the stiffening of the elastomers
at high strains, due to the finite extensibility of their polymer
chains. The corresponding strain energy density is

\begin{equation}
\hyper=-\frac{\mu}{2}\jm\ln\left(1-\frac{\ii 1-3}{\jm}\right)-\mu\ln J+\left(\frac{\bulk}{2}+\frac{\mu}{3}+\frac{\mu}{\jm}\right)\left(J-1\right)^{2},\label{eq:gentmodel}
\end{equation}
where $\jm$ models the elastomeric strain stiffening; in the limit
$\jm\rightarrow\infty$, the neo-Hookean model is recovered. Application
of Eq.$\ $\eqref{eq:wavespeeds} to the Gent model \eqref{eq:gentmodel}
provides the velocities 
\begin{equation}
c_{\pm}^{2}=\frac{a_{3}}{2\refden}+\frac{\mu}{2\refden a_{1}^{2}}+\frac{\left(a_{1}^{2}+Q^{2}+\jm\right)\mu\jm}{2\refden a_{2}^{2}}\pm\frac{\sqrt{48a_{1}^{6}\mu^{2}Q^{2}J_{m}^{2}+\left(a_{1}^{2}a_{2}^{2}a_{3}+2a_{1}^{4}\mu J_{m}-2a_{1}^{2}Q^{2}\mu\jm+a_{2}^{2}\mu\right){}^{2}}}{2\sqrt{3}\refden a_{2}^{2}a_{1}^{2}},\label{eq:cmgent-1}
\end{equation}
where $a_{1}=M+1,\ a_{2}=M^{2}+2M+Q^{2}-\jm$, and $a_{3}=\kappa-\frac{2\mu}{3}-\frac{2\mu}{\jm}$.
Notably, both $\pressurewavespeed$ and $\shearwavespeed$ depend
nonlinearly on $\shearstrain$ and $\pressurestrain$. These velocities,
in turn, yield nonlinear differential equations (\ref{eq:shearRegionEq}-\ref{eq:pressureRegionEq})
for $\shearstrain\left(\pressurestrain\right)$ and $\pressurestrain\left(\shearstrain\right)$,
which cannot be solved analytically. To investigate the dependency
of the velocities on the parameters of the problem, we use numerical
solutions obtained by the Runge-Kutta solver of the software Wolfram
Mathematica 11.3 \citet{Mathematica}

\floatsetup[figure]{style=plain,subcapbesideposition=top}
\begin{figure}[!t]
\begin{raggedright}
\centering\sidesubfloat[]{\includegraphics[scale=0.3]{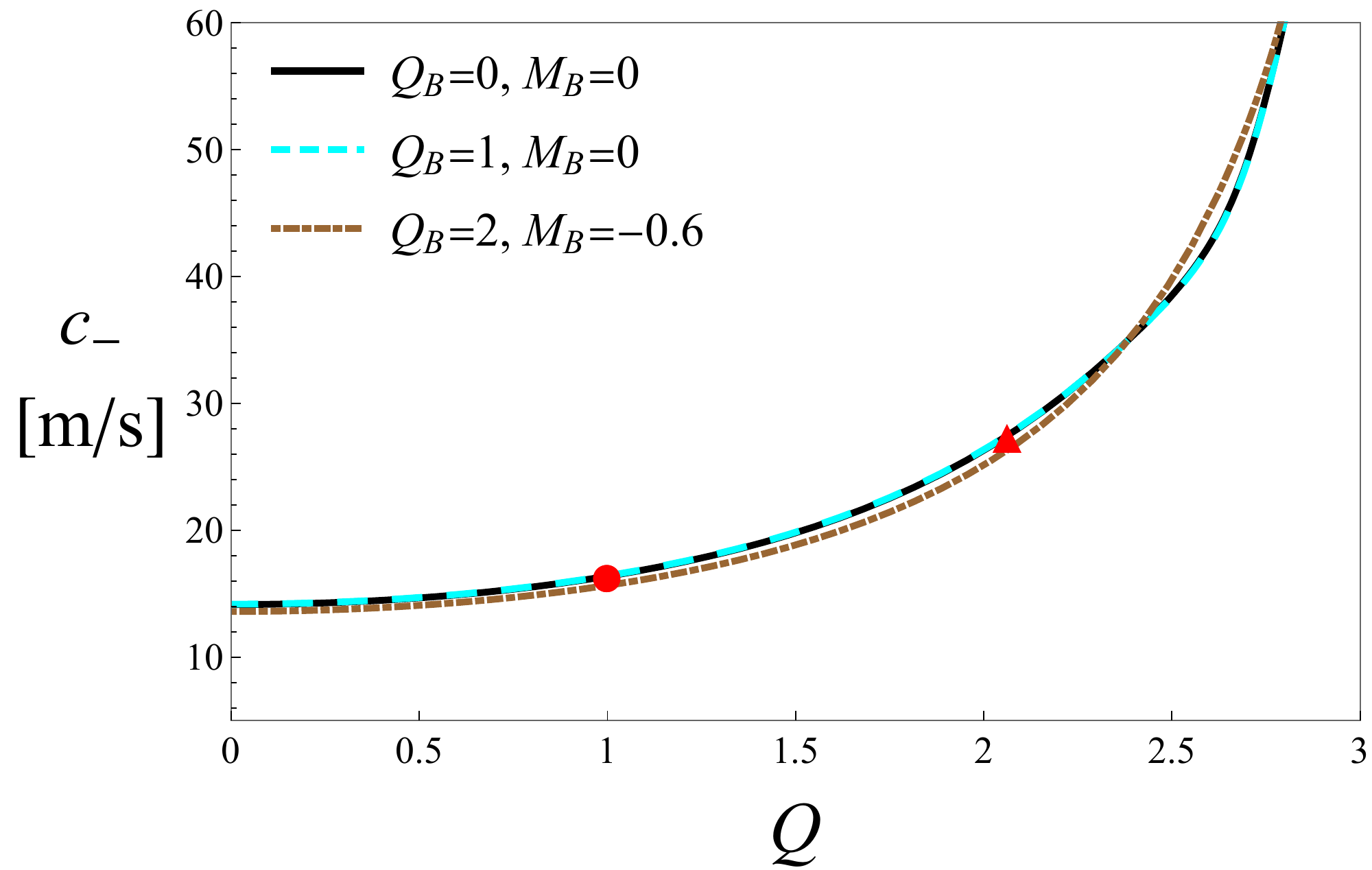}}  $\;\;\;$\sidesubfloat[]{\includegraphics[scale=0.3]{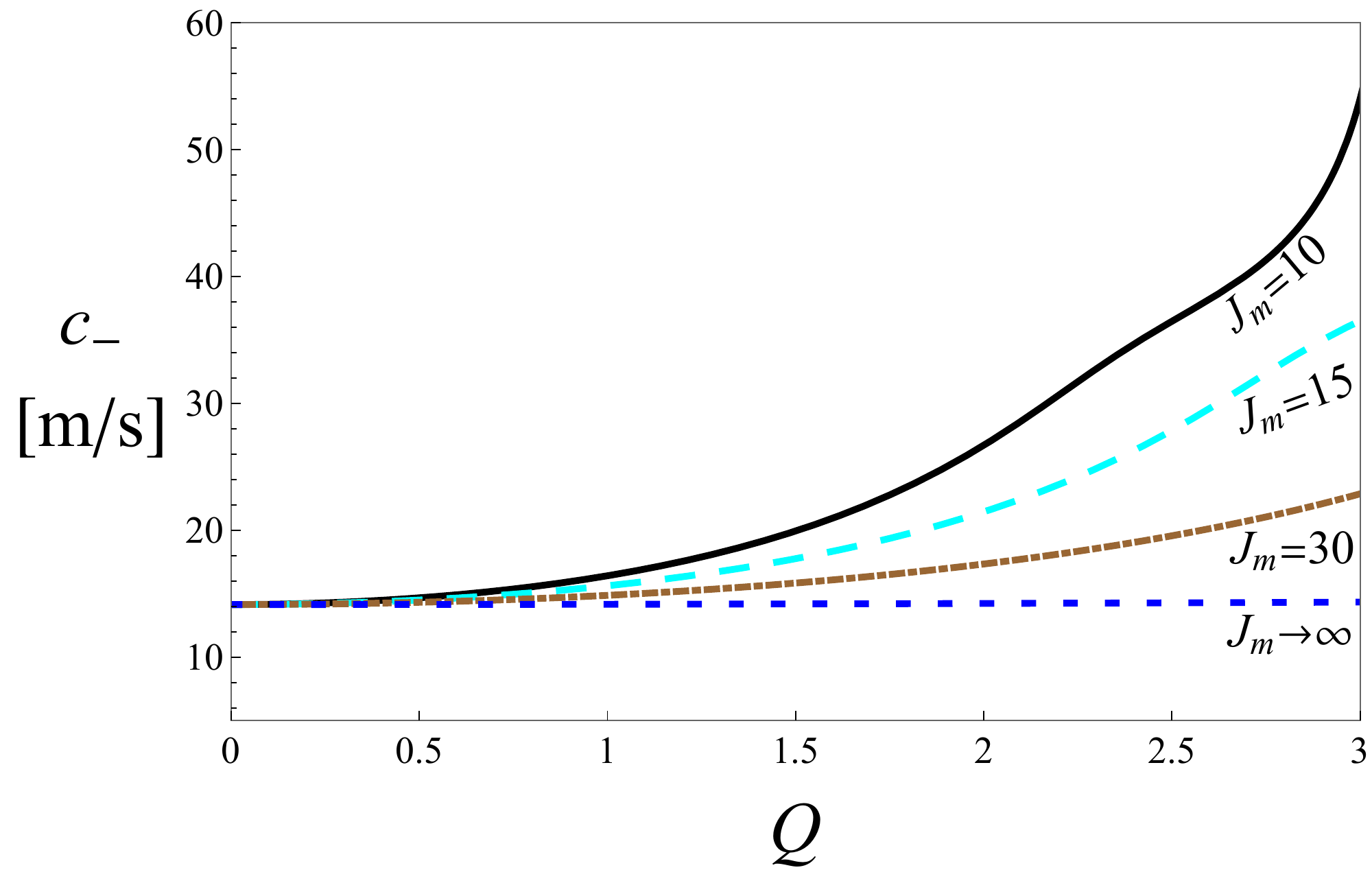}}\\
\centering\sidesubfloat[]{\includegraphics[scale=0.3]{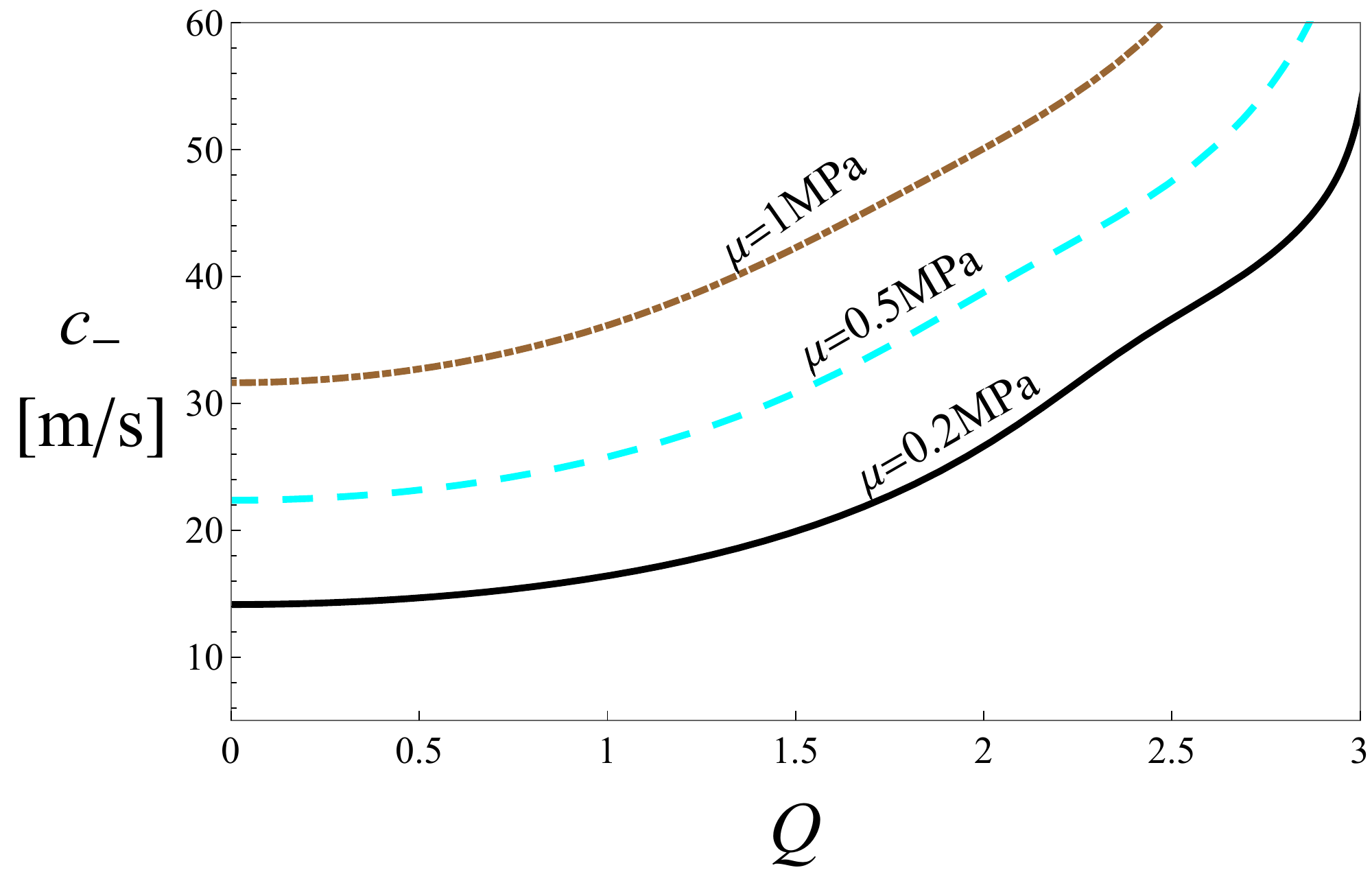}}  $\;\;\;$\sidesubfloat[]{\includegraphics[scale=0.3]{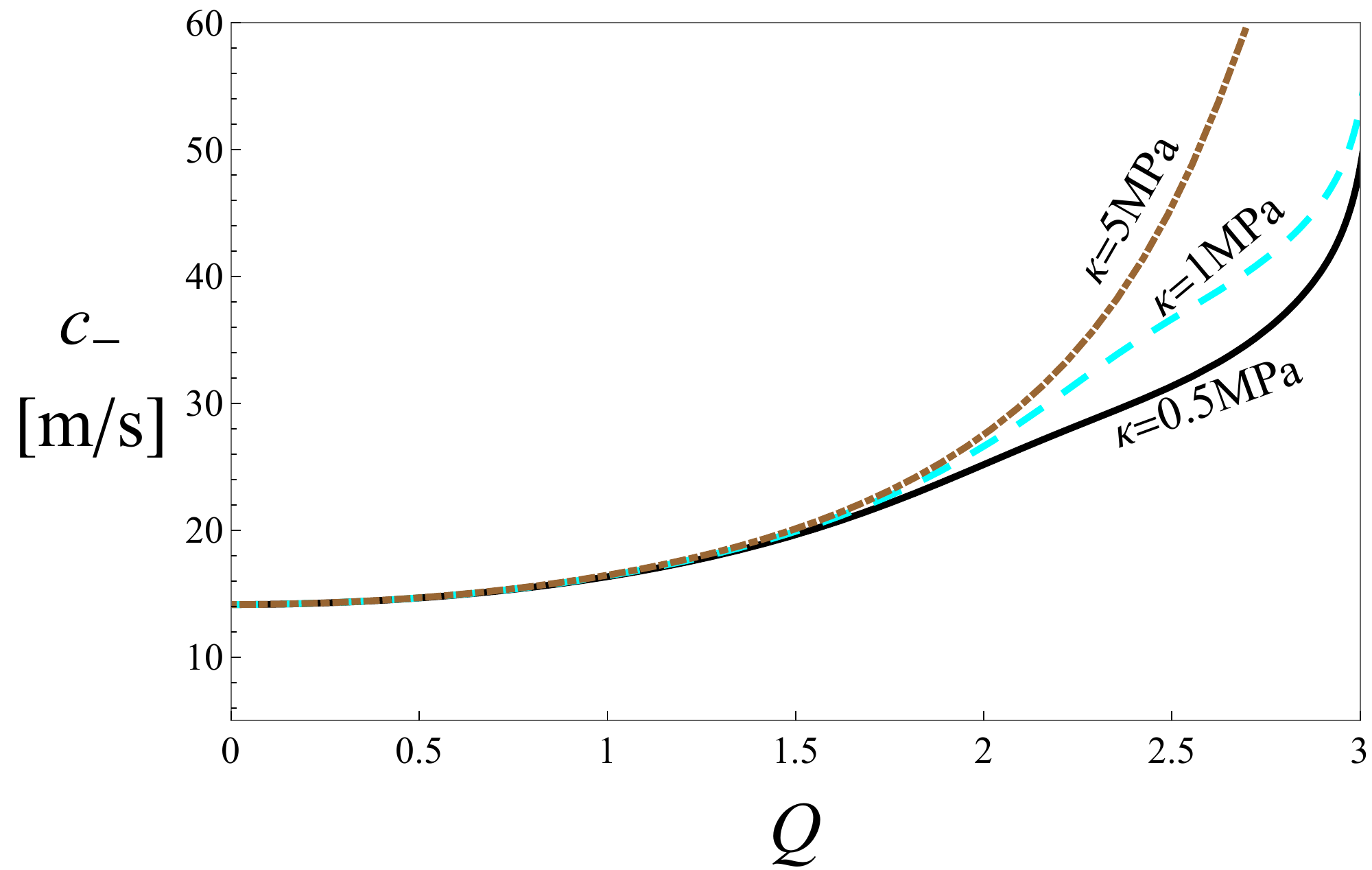}}
\par\end{raggedright}
\caption{\label{fig:Gentwavespeeds}The quasi-shear wave velocity $\protect\shearwavespeed$
as function of $\protect\shearstrain$, for the Gent material at the
boundary conditions $\protect\bshear=0,\protect\bpressure=0$, material
parameters \eqref{eq:parameters} and $\mathit{\protect\jm}=10.$
Each panel shows $\protect\shearwavespeed$ for different values of
the (a) boundary conditions: $\protect\bshear=0,\protect\bpressure=0$
(solid black), $\protect\bshear=1,\protect\bpressure=0$ (dashed cyan)
and $\protect\bshear=2,\protect\bpressure=-0.6$ (dot-dashed brown);
(b) locking parameter: $\protect\jm=10$ (solid black), $25$ (dashed
cyan), $30$ (dot-dashed brown) and $\infty$ (blue dashed); (c) shear
modulus: $\mu=0.2\mathrm{MPa}$ (solid black), $0.5\mathrm{MPa}$
(dashed cyan) and $1\mathrm{MPa}$ (dot-dashed brown); (d) bulk modulus:
$\kappa=0.5\mathrm{MPa}$ (solid black), $1\mathrm{MPa}$ (dashed
cyan) and $5\mathrm{MPa}$ (dot-dashed brown).}
\end{figure}
We begin with the study of $\shearwavespeed$ in Fig.$\ $\ref{fig:Gentwavespeeds}.
In Panel \ref{fig:Gentwavespeeds}(a) we evaluate $\shearwavespeed$
as a function of $\shearstrain$ for the parameters in Eq.$\ $\eqref{eq:parameters}
and $\jm=10$, at different impacts. Specifically, we evaluate $\shearwavespeed$
for the boundary conditions $\bshear=0,\bpressure=0$ (solid black),
$\bshear=1,\bpressure=0$ (dashed blue) and $\bshear=2,\bpressure=-0.6$
(dot-dashed brown). We observe that the boundary conditions have little
effect on the curve of $\shearwavespeed$, which is a monotonically
increasing function of $\shearstrain$. As we will demonstrate in
panels \ref{fig:Gentwavespeeds}(b)-\ref{fig:Gentwavespeeds}(d),
this monotonicity is independent of the specific material parameters.
Therefore, shear impact \emph{loading} results in shock, while shear
impact \emph{unloading} results in a smooth wave. We recall that this
phenomenon cannot be captured by the neo-Hookean model, where shear
shock cannot evolve.\floatsetup[figure]{style=plain,subcapbesideposition=top}
\begin{figure}[!t]
\begin{raggedright}
\centering\sidesubfloat[]{\includegraphics[scale=0.3]{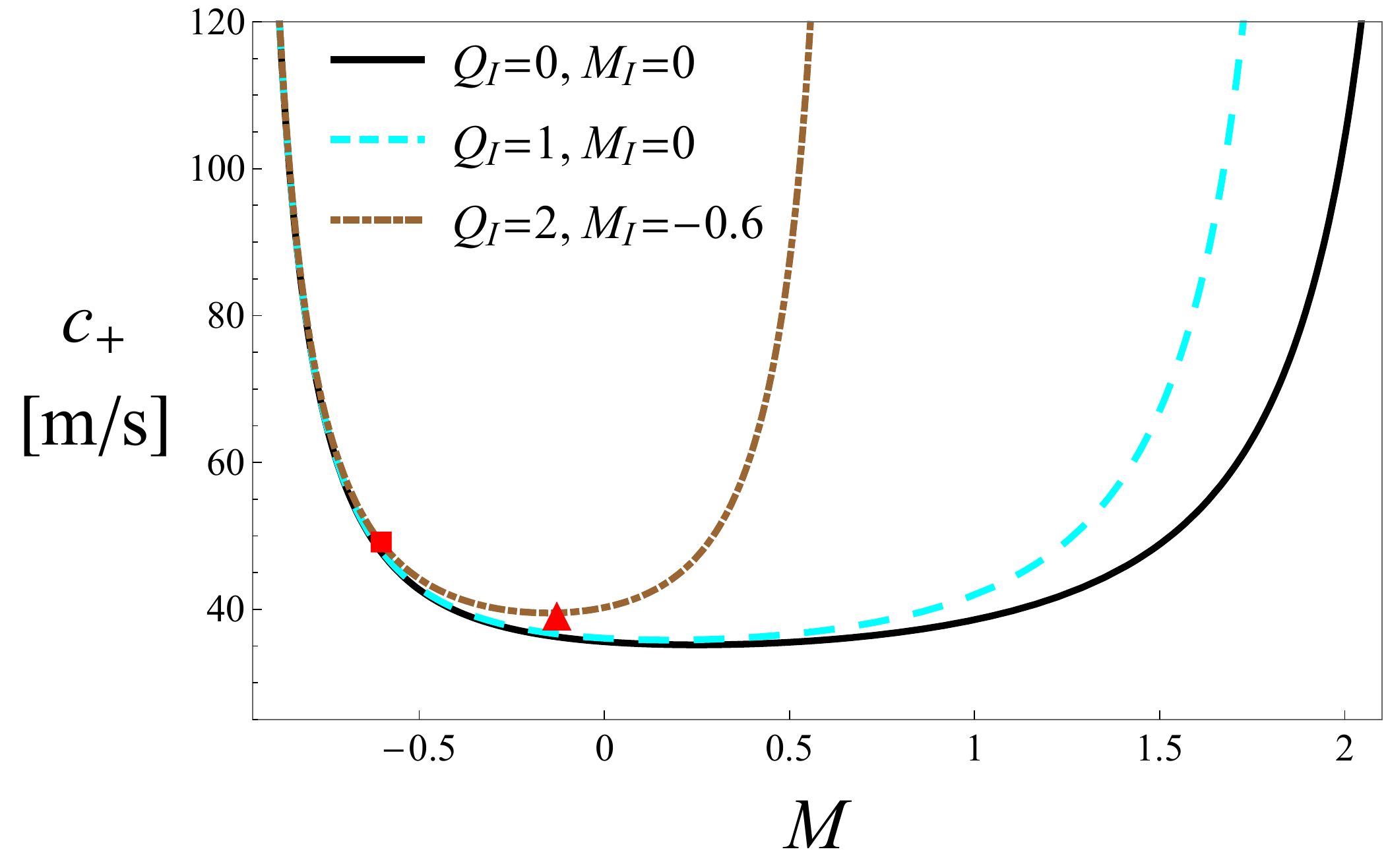}}  $\;\;\;$\sidesubfloat[]{\includegraphics[scale=0.3]{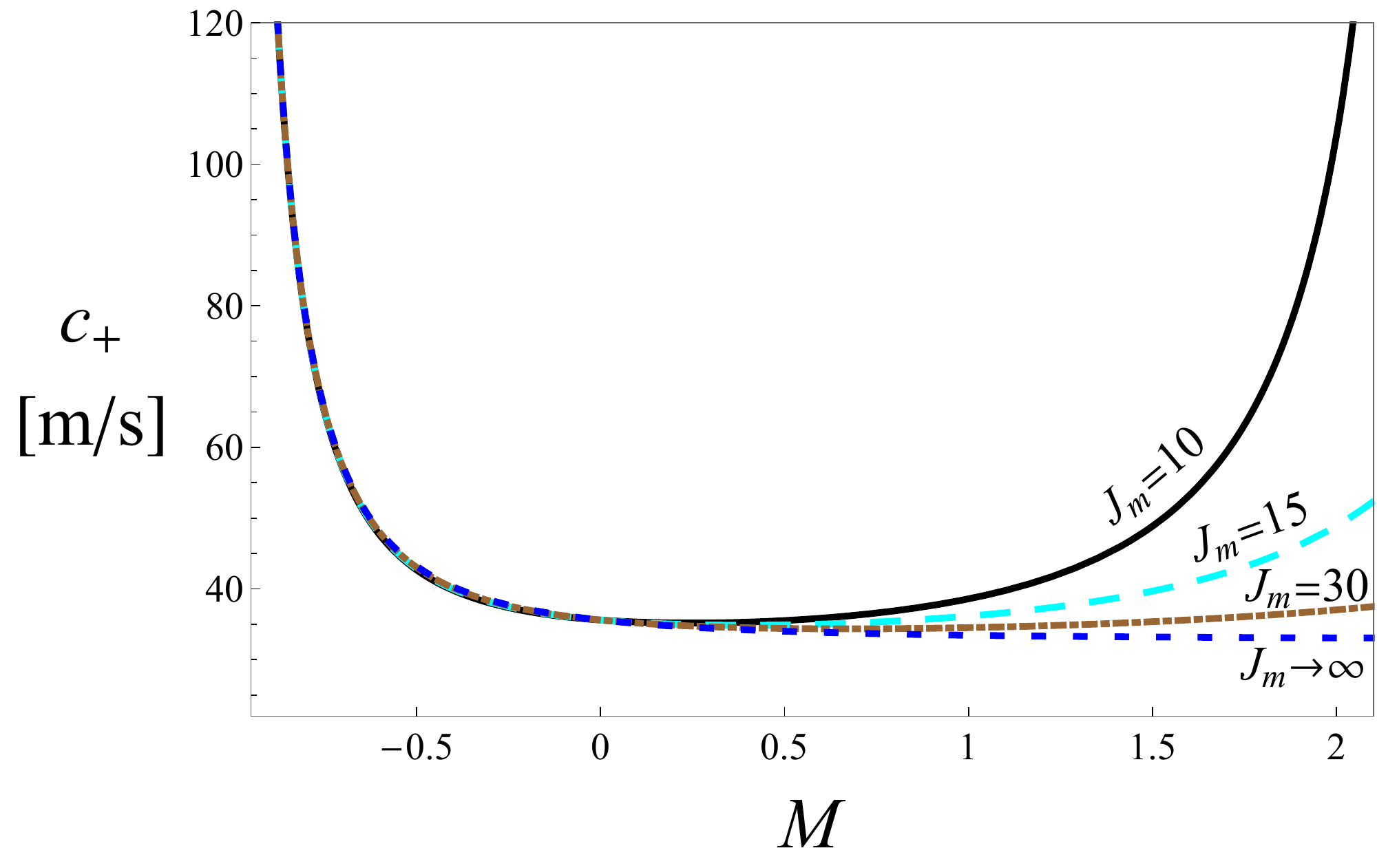}}\\
\centering\sidesubfloat[]{\includegraphics[scale=0.3]{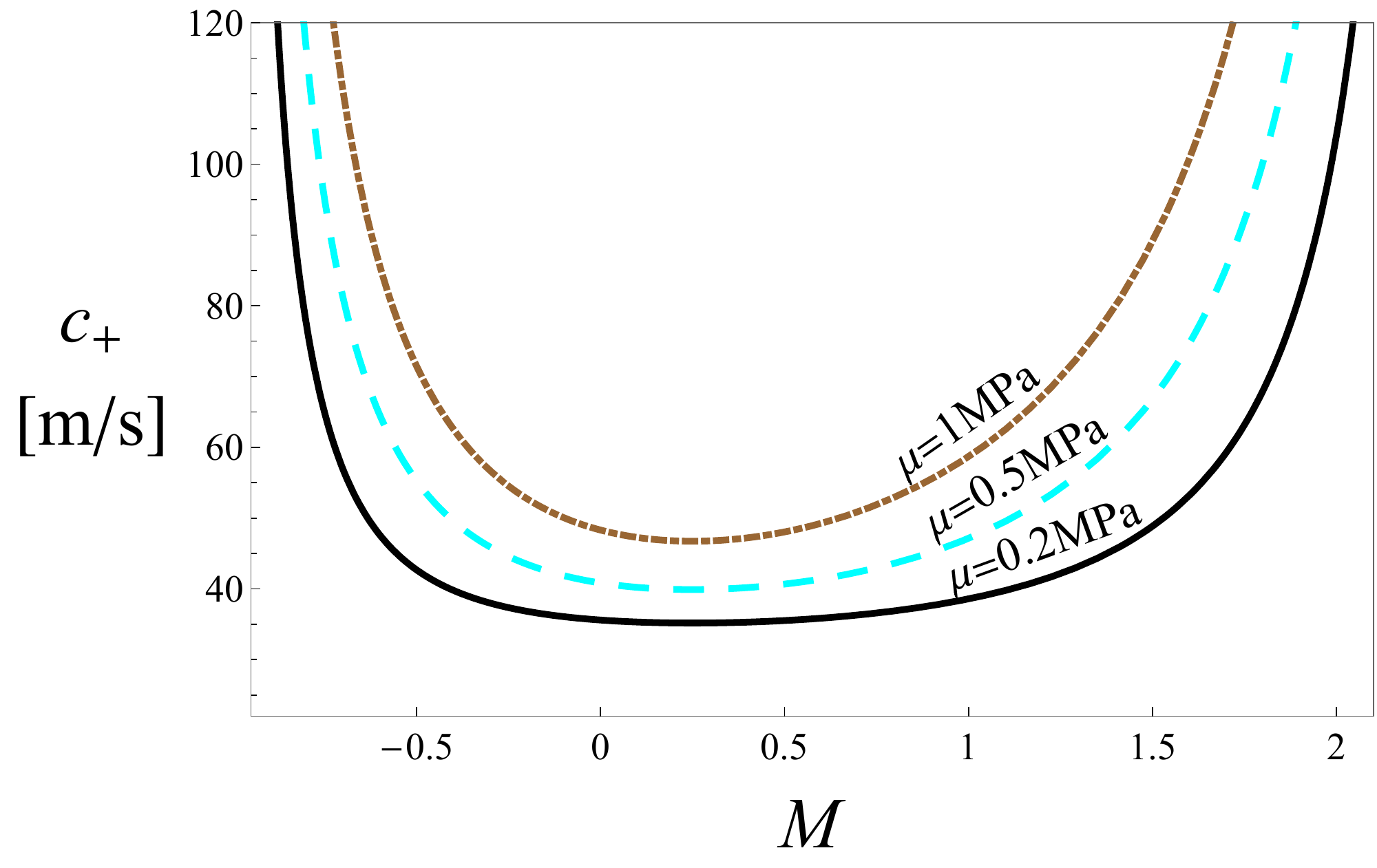}}  $\;\;\;$\sidesubfloat[]{\includegraphics[scale=0.3]{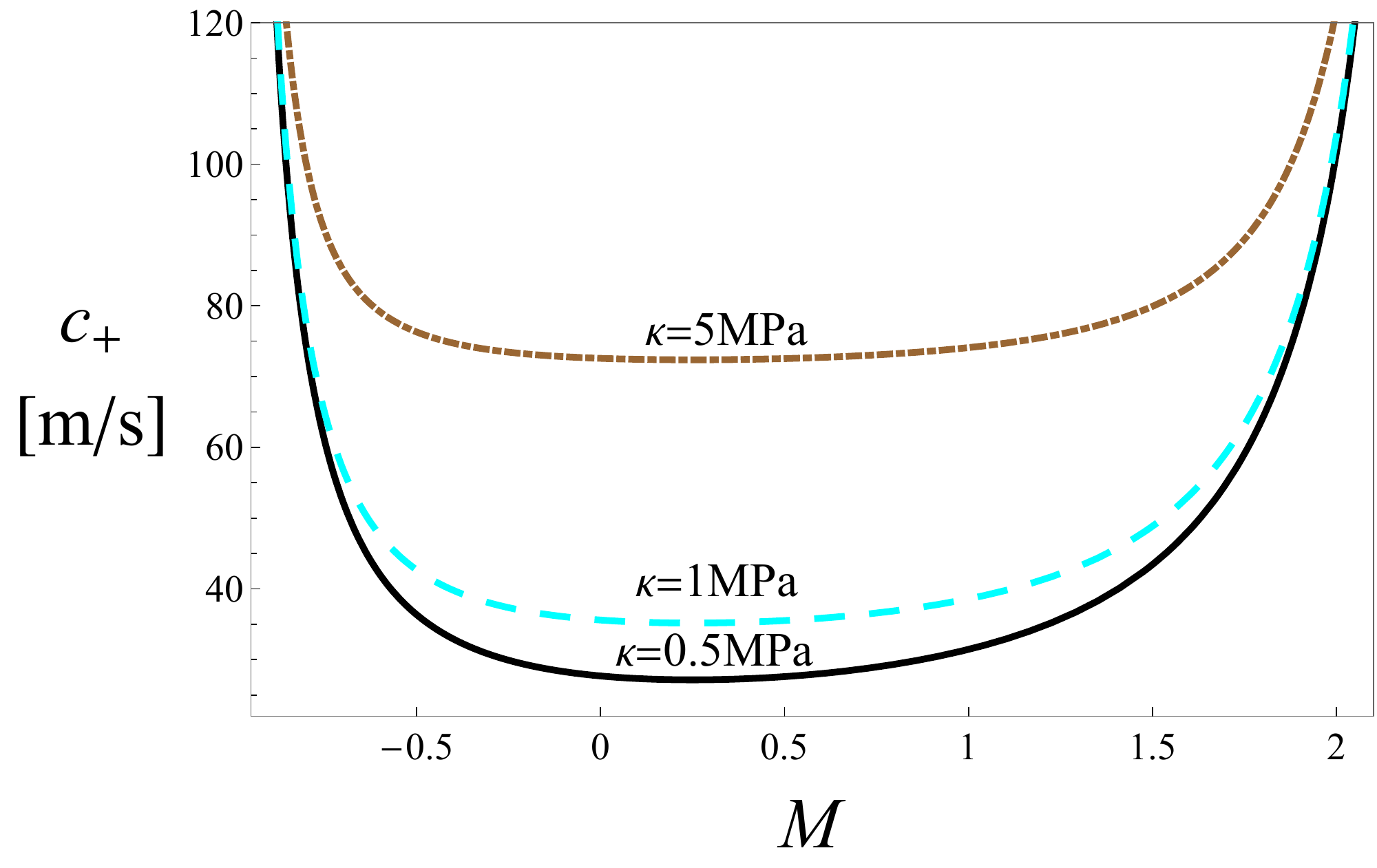}}
\par\end{raggedright}
\caption{\label{fig:GentwavespeedsPRESSURE}The pressure wave velocity $\protect\pressurewavespeed$
as function of $\protect\pressurestrain$, for the Gent material at
the boundary conditions $\protect\bshear=0,\protect\bpressure=0$,
material parameters \eqref{eq:parameters} and $\mathit{\protect\jm}=10.$
Each panel shows $\protect\pressurewavespeed$ for different values
of the (a) boundary conditions: $\protect\bshear=0,\protect\bpressure=0$
(solid black), $\protect\bshear=1,\protect\bpressure=0$ (dashed cyan)
and $\protect\bshear=2,\protect\bpressure=-0.6$ (dot-dashed brown);
(b) locking parameter: $\protect\jm=10$ (solid black), $25$ (dashed
cyan), $30$ (dot-dashed brown) and $\infty$ (dashed blue); (c) shear
modulus: $\mu=0.2\mathrm{MPa}$ (solid black), $0.5\mathrm{MPa}$
(dashed cyan) and $1\mathrm{MPa}$ (dot-dashed brown); (d) bulk modulus:
$\kappa=0.5\mathrm{MPa}$ (solid black), $1\mathrm{MPa}$ (dashed
cyan) and $5\mathrm{MPa}$ (dot-dashed brown).}
\end{figure}
Panel \ref{fig:Gentwavespeeds}(b) shows $\shearwavespeed\left(\shearstrain\right)$
for $\bshear=0,\ \bpressure=0$, and different values of the locking
parameter $\jm$. Specifically, the solid black, dashed cyan, dot-dashed
brown, and dashed blue curves correspond to $\jm=10,15,30$ and $\jm\rightarrow\infty$,
respectively. We observe that the velocity decreases as $\jm$ increases,
while its monotonicity in $\shearstrain$ is maintained. In the limit
$\jm\rightarrow\infty$ the velocity is independent of $\shearstrain$,
thereby recovering the neo-Hookean response, as it should.

Panel \ref{fig:Gentwavespeeds}(c) shows $\shearwavespeed\left(\shearstrain\right)$
for $\bshear=0,\ \bpressure=0$, and different values of the shear
modulus $\mu$. Specifically, the solid black, dashed cyan and dot-dashed
brown curves correspond to $\mu=0.2\mathrm{MPa},0.5\mathrm{MPa}$
and $1\mathrm{MPa}$, respectively. The velocity is greater for higher
values of $\mu$, as expected. Here again, the monotonicity in $\shearstrain$
is kept.

Panel \ref{fig:Gentwavespeeds}(d) shows $\shearwavespeed\left(\shearstrain\right)$
for $\bshear=0,\ \bpressure=0$, and different values of the bulk
modulus $\kappa$. Specifically, the solid black, dashed cyan and
dot-dashed brown curves correspond to $\kappa=0.5\mathrm{MPa},1\mathrm{MPa}$
and $5\mathrm{MPa}$, respectively. At small amounts of shear ($\shearstrain\leq1.5$),
the velocity is independent of $\kappa$. When the material is severely
sheared ($\shearstrain>1.5$), the coupling between the modes becomes
substantial, and the quasi-shear velocity is higher in materials with
greater bulk modulus. Again, independently of $\kappa$, the velocity
is a monotonically increasing function of $\shearstrain$.

We analyze next the quasi-pressure wave velocity $\pressurewavespeed$
in Fig.$\ $\ref{fig:GentwavespeedsPRESSURE}. Panel \ref{fig:GentwavespeedsPRESSURE}(a)
shows $\pressurewavespeed$ as function of $\pressurestrain$ for
the initial conditions $\ishear=0,\text{\ensuremath{\ipressure}}=0$
(solid black), $\ishear=1,\text{\ensuremath{\ipressure}}=0$ (dashed
cyan), and $\ishear=2,\text{\ensuremath{\ipressure}}=-0.6$ (dot-dashed
brown). Contrary to the monotonicity of $\shearwavespeed$, we observe
that beyond a critical deformation, denoted $\cpressure$, the curve
changes its trend from downward to upward. The value of $\cpressure$
decreases for greater values of $\ishear$. Since the condition for
smooth waves depends on the sign of $\frac{\mathrm{d}\pressurewavespeed}{\mathrm{d}\pressurestrain}$
and the location of $\upressure,\ipressure$ and $\cpressure$, we
can deduce if shocks emerge from this diagram, as we will demonstrate
later.

Panel \ref{fig:GentwavespeedsPRESSURE}(b) shows $\pressurewavespeed\left(\pressurestrain\right)$
for $\ishear=0,\text{\ensuremath{\ipressure}}=0$, and different values
of the locking parameter $\jm$. The legend is identical to the legend
in panel \ref{fig:Gentwavespeeds}(b). We observe that the slope beyond
$\cpressure$ decreases as $\jm$ increases, until it vanishes in
the limit $\jm\rightarrow\infty$, thereby recovering the neo-Hookean
response, as it should.

Panel \ref{fig:GentwavespeedsPRESSURE}(c) shows $\pressurewavespeed\left(\pressurestrain\right)$
for $\ishear=0,\text{\ensuremath{\ipressure}}=0$, and different values
of the shear modulus $\mu$. The legend is identical to the legend
in panel \ref{fig:Gentwavespeeds}(c). The velocity is higher for
greater values of $\mu$, while the value of $\cpressure$ remains
unchanged.

Panel \ref{fig:GentwavespeedsPRESSURE}(d) shows $\pressurewavespeed\left(\pressurestrain\right)$
for $\ishear=0,\text{\ensuremath{\ipressure}}=0$, and different values
of the bulk modulus $\kappa$. The legend is identical to the legend
in panel \ref{fig:Gentwavespeeds}(d). As expected, the velocity is
higher when $\kappa$ is greater, where the difference decreases as
the strain increases. We observe that the value of $\cpressure$ is
independent of $\kappa$.

\section{Combined shear and axial impact of unstrained materials\label{sec:atRest}}

We provide next a more comprehensive analysis of the effect of the
boundary conditions $\bshear$ and $\bpressure$ on waves in unstrained
materials ($\ipressure=0,\;\ishear=0$), starting with quasi-shear
waves. We recall that shear waves in neo-Hookean materials always
propagate as smooth waves, since $\shearwavespeed$ is constant. To
analyze quasi-shear waves in Gent materials, we utilize Fig.$\ $\ref{fig:Gentwavespeeds}(a),
to deduce that these waves will always propagate as shocks, since
$\frac{\mathrm{d}\shearwavespeed}{\mathrm{d}Q}>0$ for any $Q_{B}>0=Q_{I}$.

\floatsetup[figure]{style=plain,subcapbesideposition=top}
\begin{figure}[!t]
\begin{raggedright}
\centering\sidesubfloat[]{\includegraphics[scale=0.27]{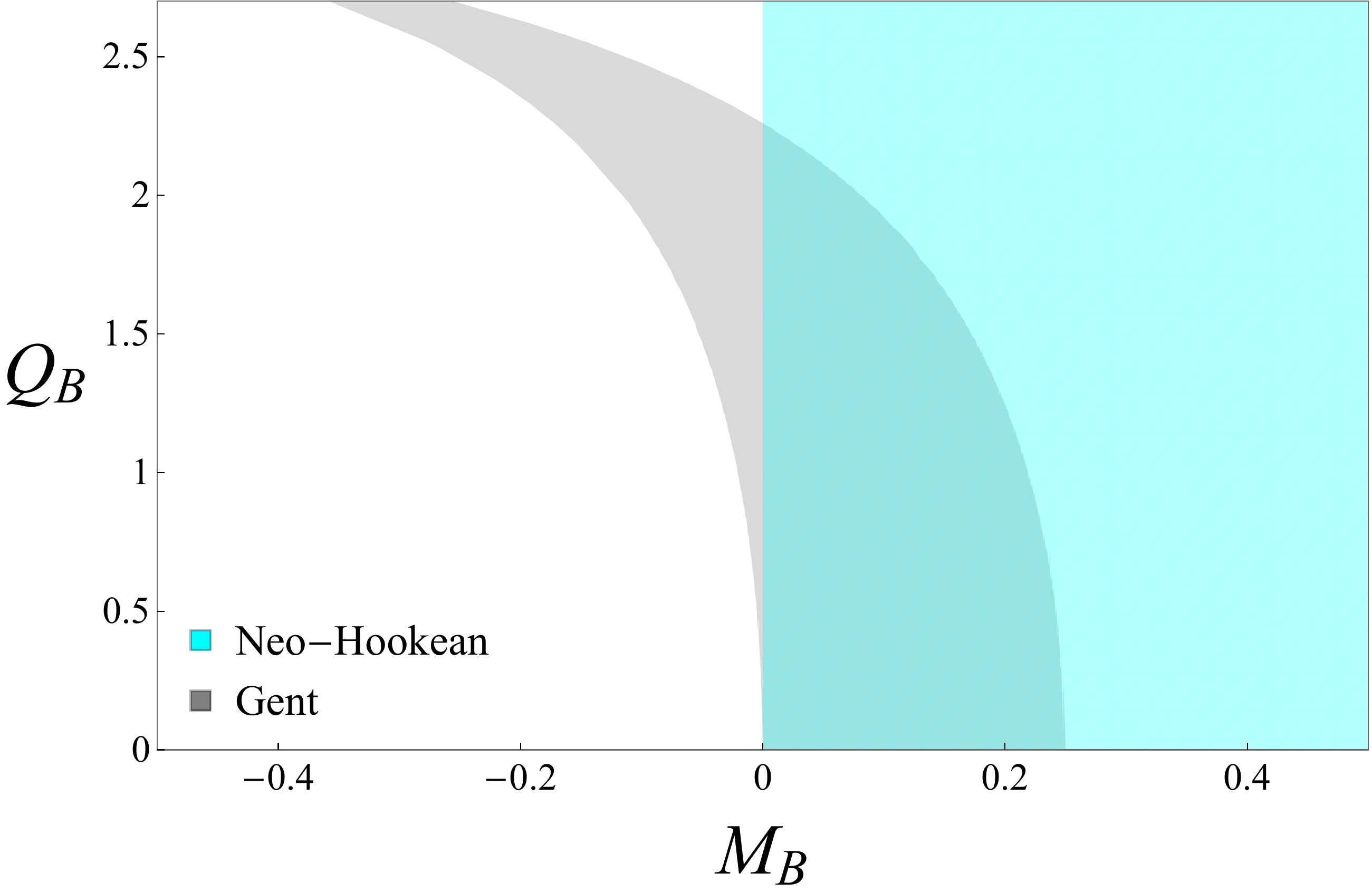}}  $\;\;\;$\sidesubfloat[]{\includegraphics[scale=0.3]{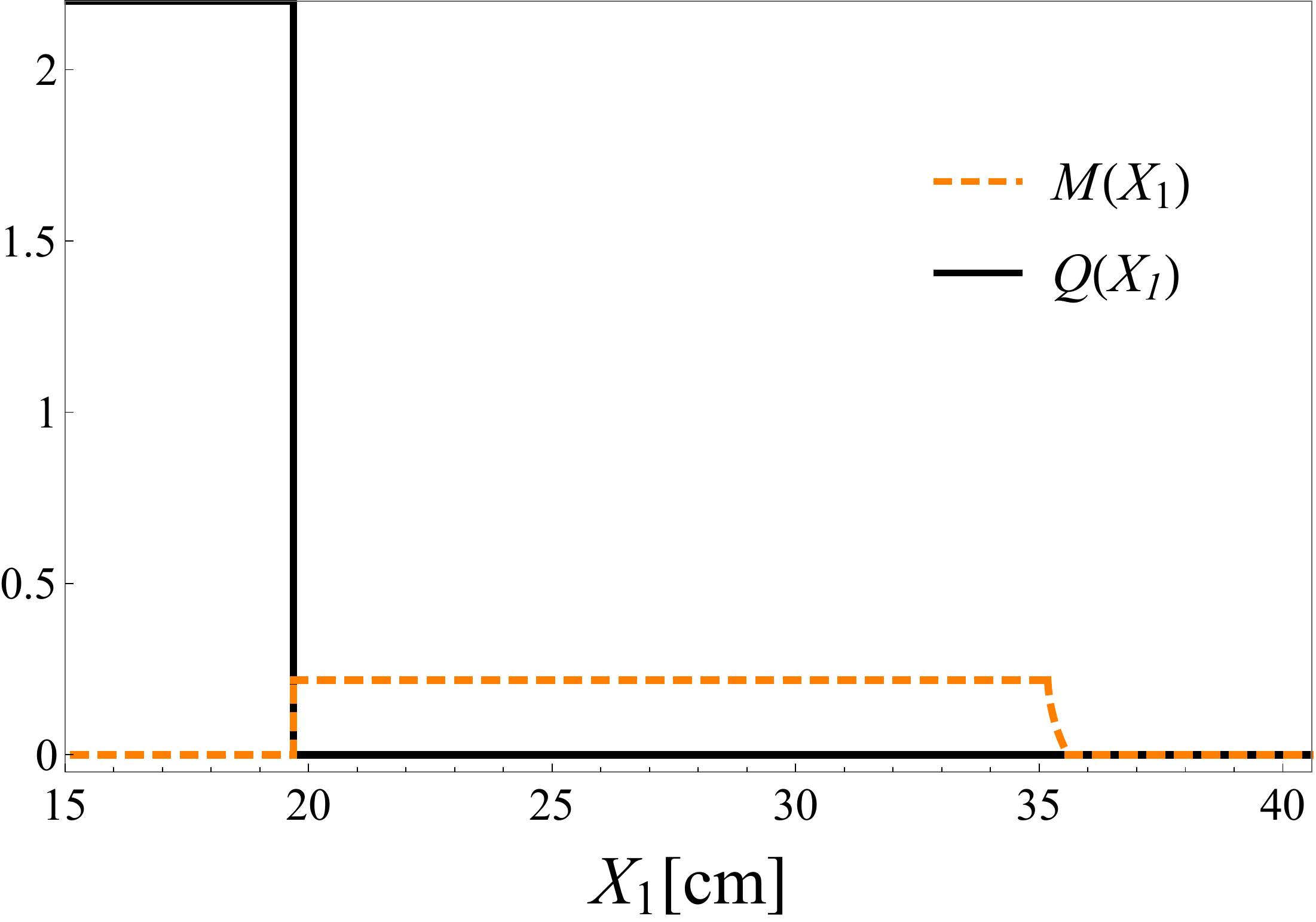}}
\par\end{raggedright}
\caption{(a) Impacts that yield smooth pressure waves in unstrained materials
as regions in the ($\protect\bpressure,\protect\bshear$)-plane. Cyan
and Gray denote neo-Hookean and Gent materials, respectively. The
material parameters are given in Eq.$\ $\eqref{eq:parameters}, where
for the Gent material we also set $\protect\jm=10$. (b) Shock-smooth
wave solutions for an unstrained Gent material, for the boundary conditions
$\protect\bpressure=0,\protect\bshear=2.2$. (b) the strain fields
$\protect\pressurestrain$ (dashed orange) and $\protect\shearstrain$
(solid black) from the shock-smooth wave due to the impact $\left(\protect\bpressure,\protect\bshear\right)=\left(0,2.2\right)$
at $t=10\,\mathrm{ms}$.\label{fig:characterizationLoading}}
\end{figure}
We proceed with the analysis of the effect on quasi-pressure waves,
which requires the calculation of $\upressure$. As stated in Sec.$\ $\ref{sec:neo-Hookean},
the value for neo-Hookean materials is simply $\upressure=\bpressure$,
since the waves are not coupled. Then, independently of $\bshear$
and for any $\bpressure<0$, pressure shocks emerge, since $\frac{\mathrm{d}\pressurewavespeed}{\mathrm{d}\pressurestrain}<0$
from $\bpressure<0$ to $\ipressure=0$ (Fig.$\ $\ref{fig:a4Neo}a).
Hence, in the $\left(\bshear,\bpressure\right)$-plane we identify
loadings that result in smooth quasi-pressure waves with the half-space
$\bpressure>0$, as illustrated in Fig.$\ $\ref{fig:characterizationLoading}(a)
by the cyan color. Thus, only compressive impacts create shocks. Again
we note that this is incompatible with experiments showing tensile-induced
shocks.

The value of $\upressure$ in Gent materials depends on $\bpressure$
\emph{and} $\bshear$, due to the coupling between the fields, and
requires the use of Eq.$\ $\eqref{eq:reducedJumpCond}$_{1}$. Note
that $\ushear=0$ in Eq.$\ $\eqref{eq:reducedJumpCond}$_{1}$, since
we assume that the shear impact is slower than $\speedSU$. Smooth
quasi-pressure waves propagate when $0<\upressure<\cpressure$, since
in this interval $\frac{\mathrm{d}\pressurewavespeed}{\mathrm{d}\pressurestrain}<0$,
as illustrated in Fig.$\ $\ref{fig:GentwavespeedsPRESSURE}(a). Conversely,
quasi-pressure shocks propagate when $\cpressure<\upressure$, since
in this interval $\pressurewavespeed$ losses its monotonicity. The
combinations of $\bshear$ and $\bpressure$ that yield $0<\upressure<\cpressure$
and smooth quasi-pressure waves are illustrated in Fig.$\ $\ref{fig:characterizationLoading}(a)
on the ($\bpressure,\bshear$)-plane by the gray color. Interestingly,
in the absence of shear impact ($\bshear=0$), there exists a threshold
value of tensile impact ($\bpressure=0.25$) above which shocks emerge.
In other words, the Gent model predicts that sufficiently strong tensile
impacts induce shocks. Notably, \citet{knowles2002impact} arrived
to the same conclusion when analyzing the tensile impact of soft rods
with cubic stress-strain equation, using the concept of \emph{thermodynamic
driving force. }We observe that the threshold value of the tensile
impact for shock is lowered by applying simultaneously shear impact.
Furthermore, the application of shear impact creates a threshold value
for compressive impacts, below which shocks cannot evolve. We conclude
this part by evaluating in panel \ref{fig:characterizationLoading}(b)
the strain fields $\pressurestrain$ (dashed orange) and $\shearstrain$
(solid black) from the shock-smooth wave at the boundary conditions
$\left(\bpressure,\bshear\right)=\left(0,2.2\right)$ at $t=10\,\mathrm{ms}$\footnote{The procedure of obtaining $\upressure$ for $\left(\ipressure,\ishear,\bpressure,\bshear\right)=\left(0,0,0.2.2\right)$
is illustrated in Appendix \ref{Appendix-A}.}. We observe that the quasi-shear shock excites also axial strains
($19.5\,\mathrm{cm}<X_{1}<35.5\,\mathrm{cm}$), and the quasi-pressure
smooth wave is not accompanied with shear strains ($35\,\mathrm{cm}<X_{1}<35.5\,\mathrm{cm}$). 

\section{Combined shear and axial impact of finitely strained materials\label{sec:strained}}

We complete our study by analyzing more extensively the general case
of pre-strained materials subjected to combined impact. We recall
that for neo-Hookean materials, the field $\upressure$ is independent
of $\ipressure$ and $\ishear$, hence we focus on Gent materials.
The complexity of the corresponding analysis stems from the fact that
we do not know \emph{a priori} which kind of combination of waves
develops, since $\upressure$ and $\ushear$ are functions of all
the prescribed quantities \{$\ipressure,\ishear,\bpressure,\bshear$\},
and $\cpressure$ depends on $\ipressure$ and $\ishear$. As mentioned
in Sec.$\ $\ref{sec:Problem-statement}, we proceed by assuming that
the combination is smooth-smooth (Fig.$\ $\ref{fig:simpleAndshockWaves}a).
We solve Eqs.$\ $\eqref{eq:shearRegionEq}-\eqref{eq:pressureRegionEq}
to determine $\upressure$ and $\ushear$ and examine if our assumption
holds via Eq.$\ $\eqref{eq:simplewaverequirement}; otherwise, we
assume the combination is shock-smooth (Fig.$\ $\ref{fig:simpleAndshockWaves}b),
and solve Eqs.$\ $\eqref{eq:pressureRegionEq} and \eqref{eq:reducedJumpCond}$_{1}$
to determine $\upressure$ and $\ushear$, and examine if our solution
satisfies Eq.$\ $\eqref{eq:simplewaverequirement} and so forth,
until we find a compatible set.\floatsetup[figure]{style=plain,subcapbesideposition=top}
\begin{figure}[!t]
\begin{centering}
\centering\sidesubfloat[]{\includegraphics[scale=0.33]{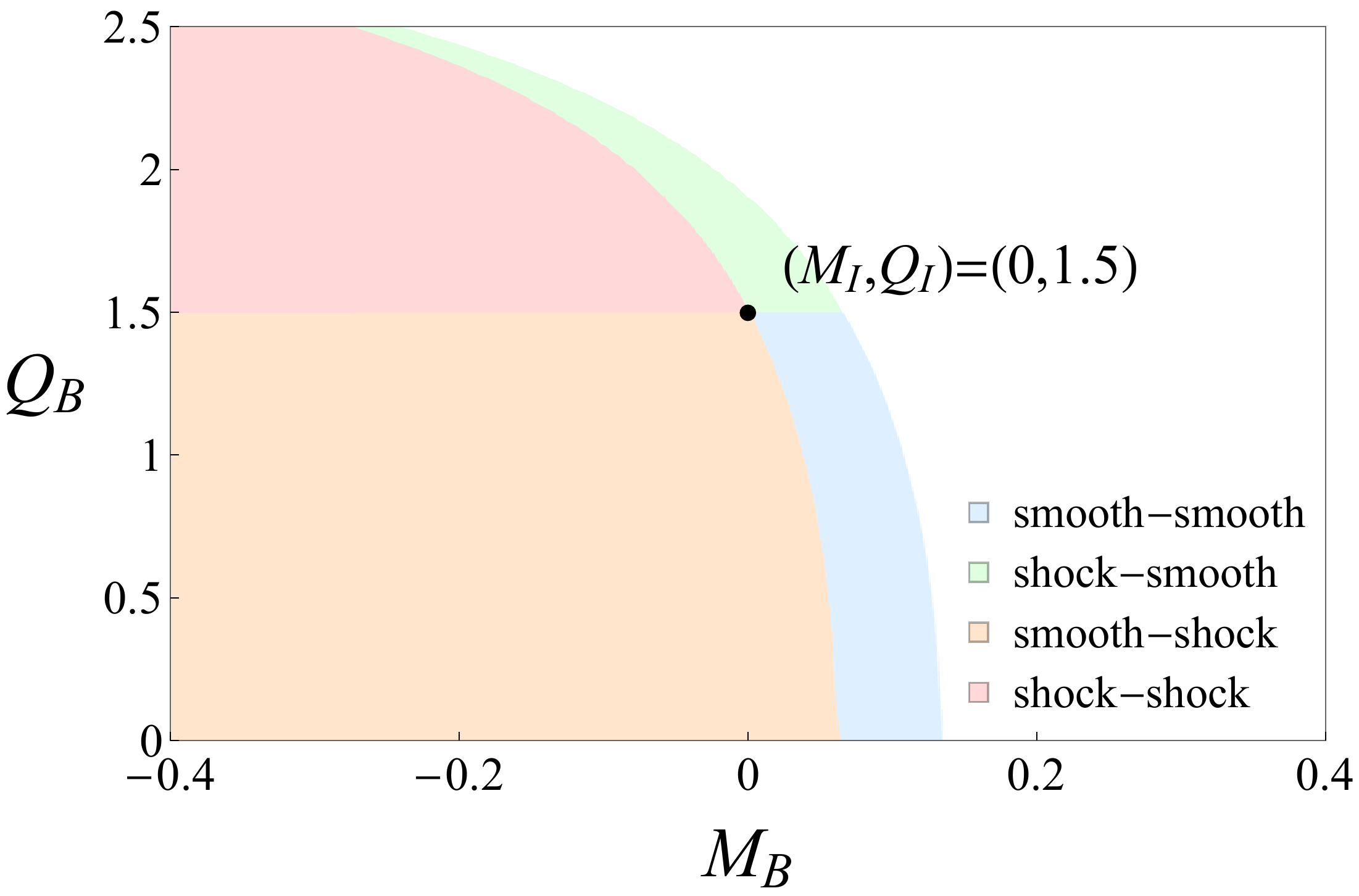}}  $\;\;\;$\sidesubfloat[]{\includegraphics[scale=0.33]{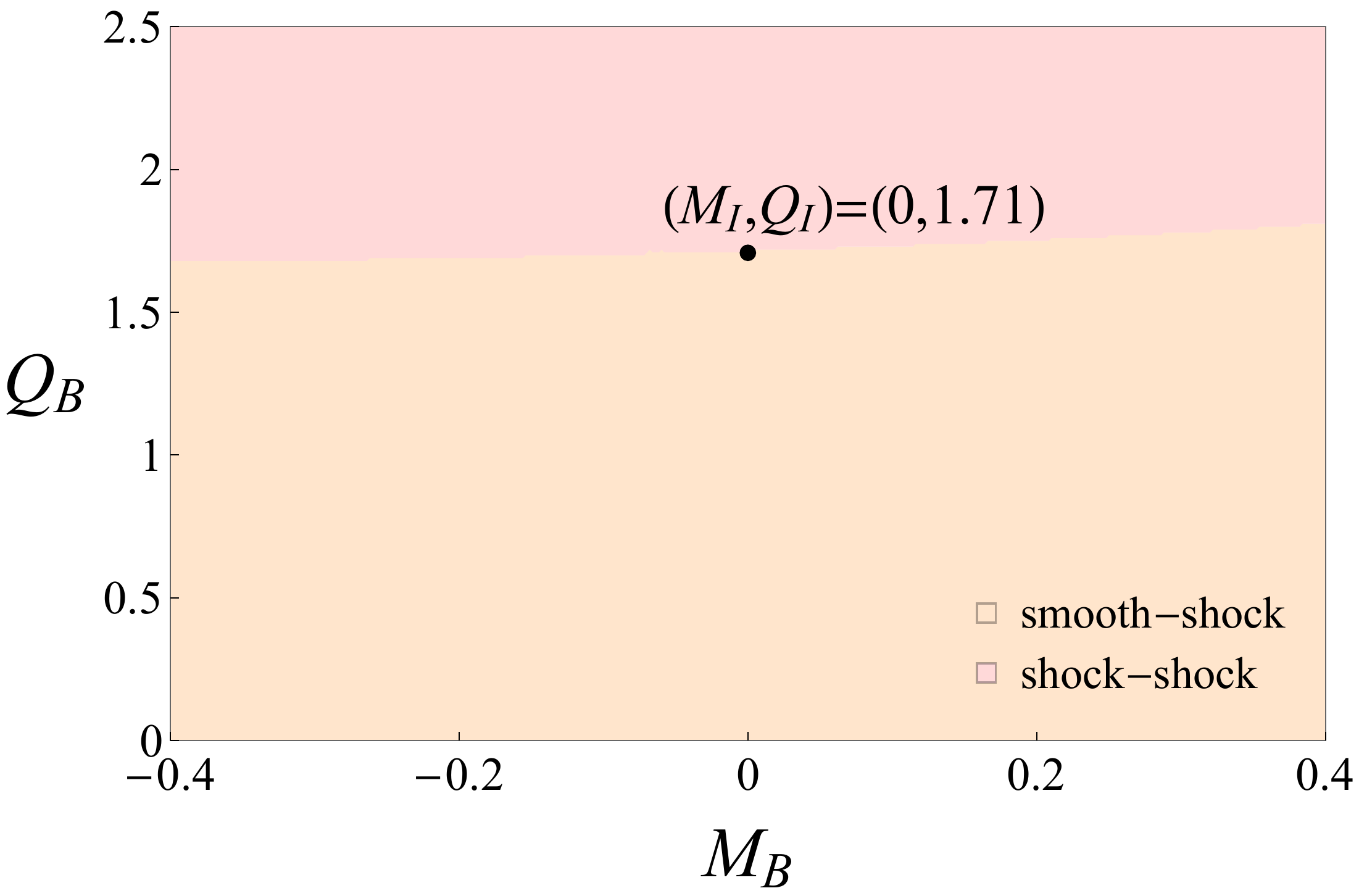}}\\
\sidesubfloat[]{\includegraphics[scale=0.35]{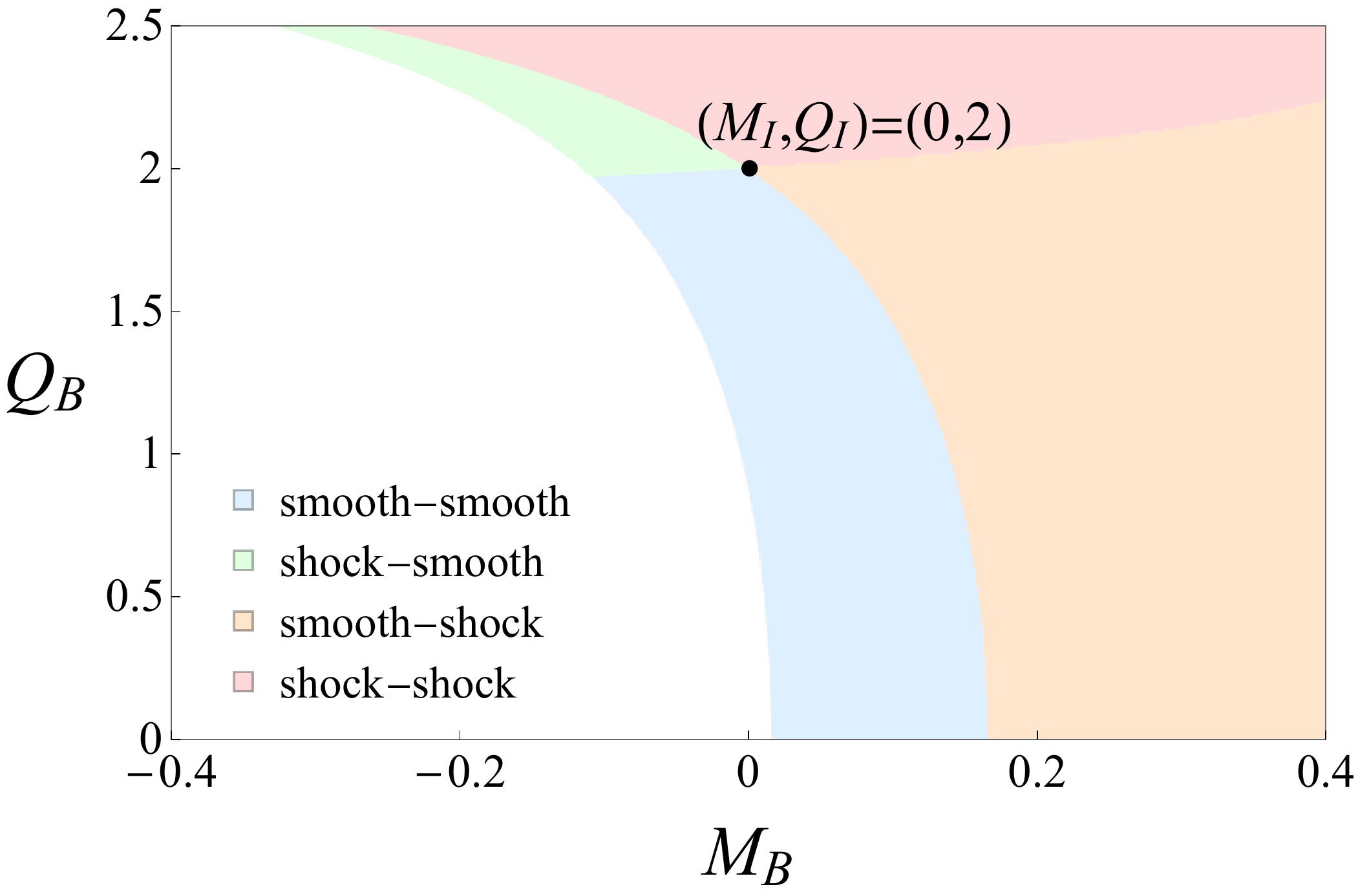}}  $\;\;\;$\sidesubfloat[]{\includegraphics[scale=0.35]{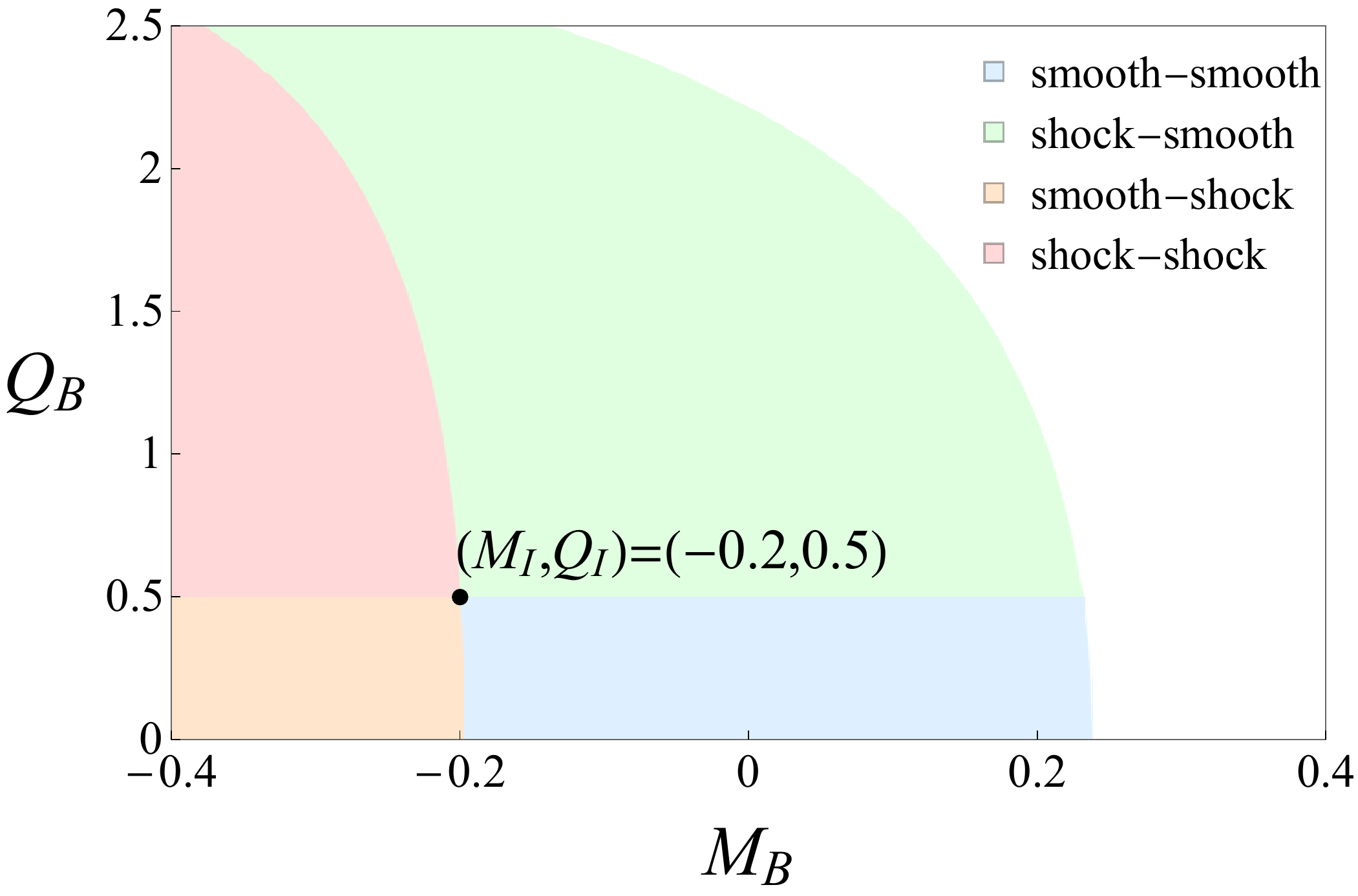}}\\
\sidesubfloat[]{\includegraphics[scale=0.33]{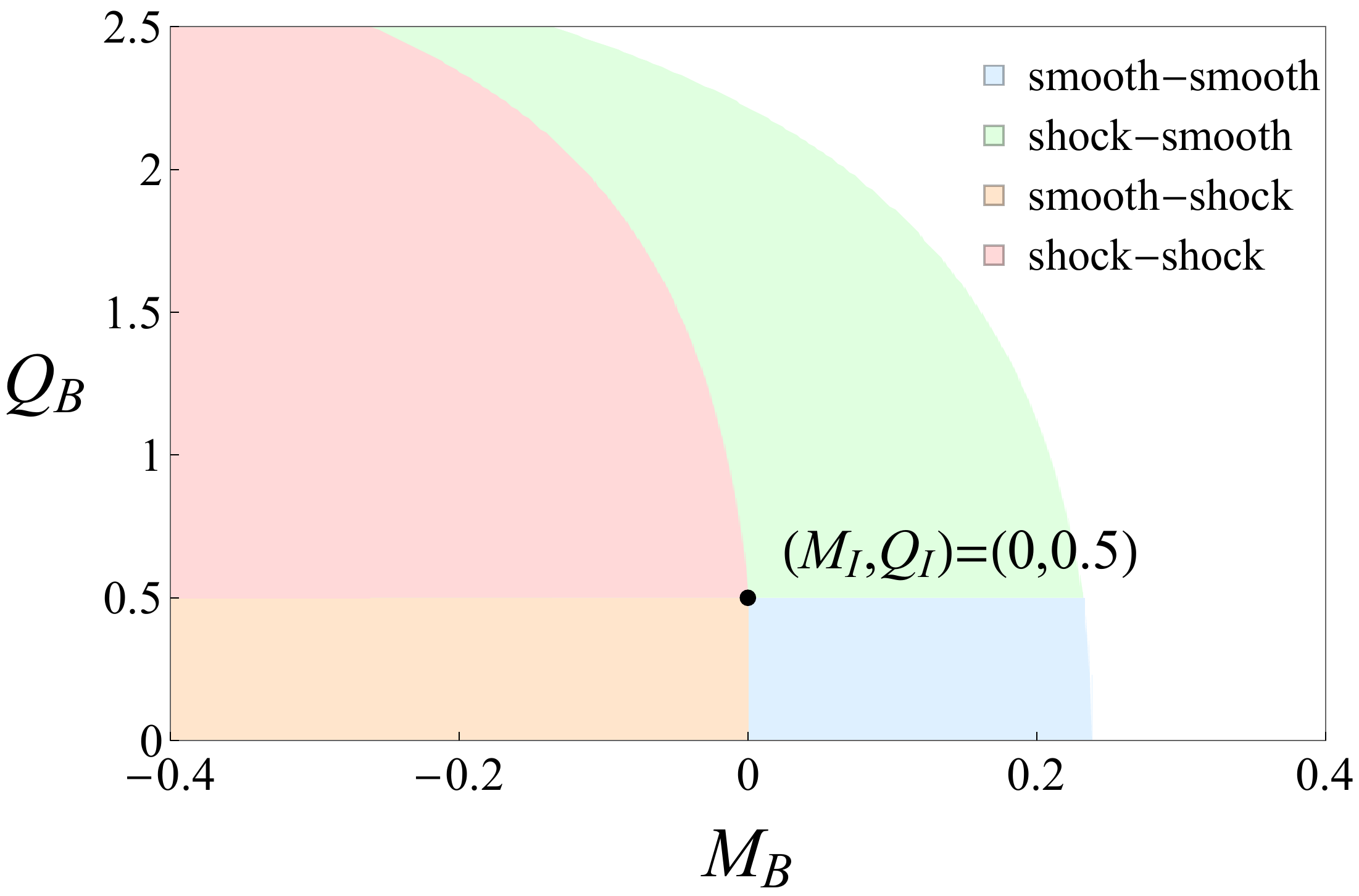}}  $\;\;\;$\sidesubfloat[]{\includegraphics[scale=0.33]{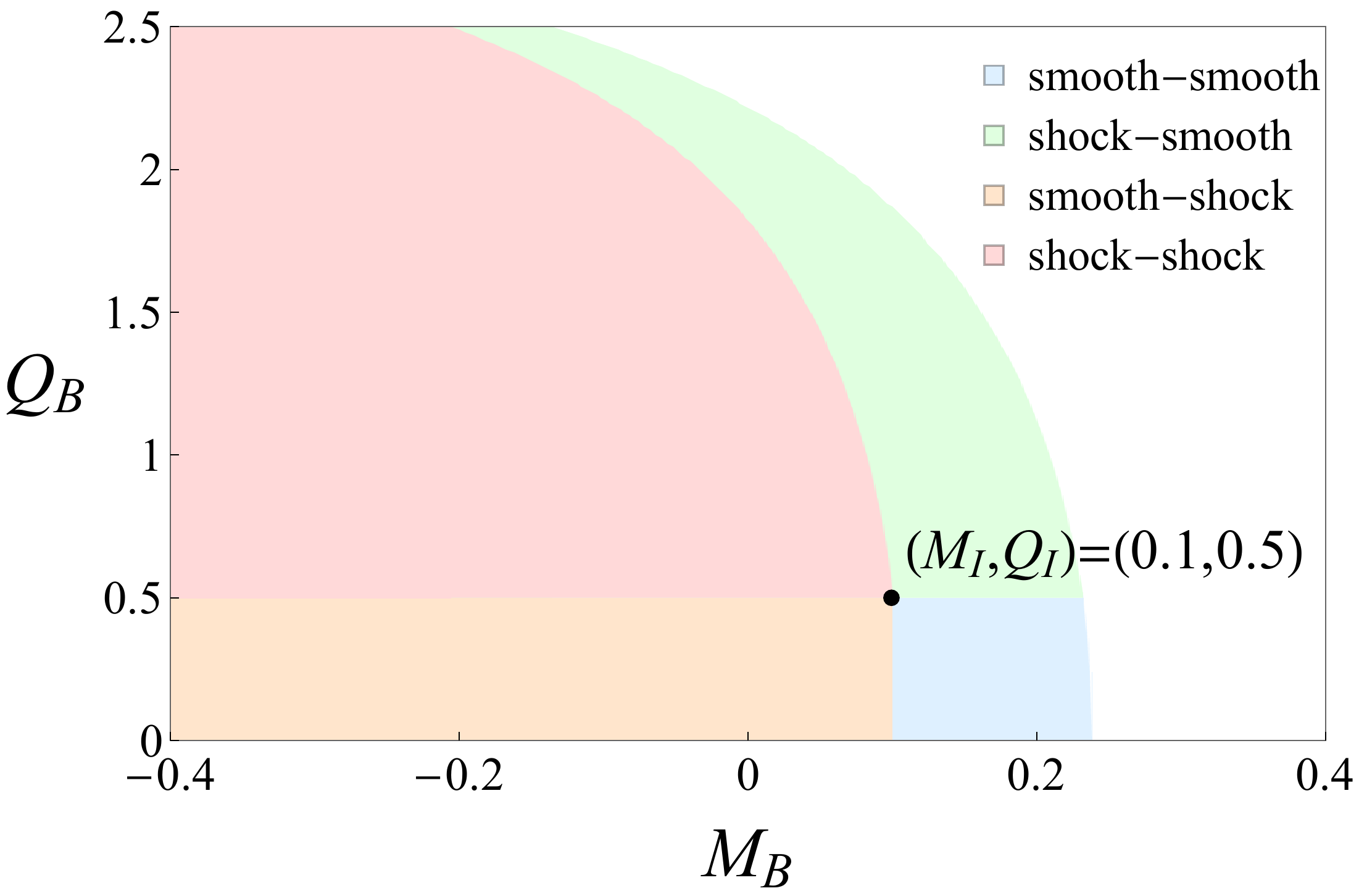}}
\par\end{centering}
\caption{Regions in the ($\protect\bpressure,\protect\bshear$)-plane that
induce smooth-smooth (blue), shock-smooth (green), smooth-shock (orange),
and shock-shock (red) waves in Gent materials with the properties
\eqref{eq:parameters} and $\protect\jm=10$. Panels (a)-(f) correspond
to $\left(\protect\ipressure,\protect\ishear\right)=\left(0,1.5\right),\ \left(0,1.71\right),\ \left(0,2\right),\ \left(-0.2,0.5\right),\ \left(0,0.5\right)$,
and $\left(0.1,0.5\right)$, respectively.\label{fig:GentMQ_general}}
\end{figure}
We illustrate in Fig.$\ $\ref{fig:GentMQ_general} the resultant
wave classification as functions of $\bpressure$ and $\bshear$ for
Gent materials with the properties \eqref{eq:parameters} and $\jm=10$.
Specifically, panels \ref{fig:GentMQ_general}(a)-\ref{fig:GentMQ_general}(d)
correspond to $\left(\ipressure,\ishear\right)=\left(0,1.5\right),\ \left(0,1.71\right),\ \left(0,2\right),\ \left(-0.2,0.5\right),\ \left(0,0.5\right)$,
and $\left(0.1,0.5\right)$, respectively. The blue, green, orange
and red regions correspond to smooth-smooth, shock-smooth, smooth-shock
and shock-shock, respectively. The white regions correspond to impacts
for shocks evolve due to loss of monotonicity of $\pressurewavespeed$,
and are outside our scope. 

We observe that by moving vertically up (resp.$\ $down), \emph{i.e.},
imparting shear loading (resp.$\ $unloading) impact, we always enter
a quasi-shear shock (resp.$\ $smooth) region\footnote{The admissibility conditions for quasi-shear waves hold also in the
white regions next to the blue regions. The admissibility conditions
for the quasi-shear waves is violated in the white regions next to
the green regions.}. During shear impact we may or may not enter a quasi-pressure shock
region, depending on the initial strain. For instance, in panel \ref{fig:GentMQ_general}(a),
shear loading impact results in smooth quasi-pressure waves, while
in panel \ref{fig:GentMQ_general}(c) it results in quasi-pressure
shocks. These panels demonstrate that the results are reversed upon
impact unloading.

The effect of axial impacts is more complex. Specifically, compressive
impact and tensile impact may or may not trigger quasi-pressure shock,
depending on the initial state. For example, panel \ref{fig:GentMQ_general}(a)
shows that only compressive impact induces quasi-pressure shock when
the pre-strain is $\ishear=1.5$; when the initial shear is increased
to 2 (panel \ref{fig:GentMQ_general}c), the trend is reversed, \emph{i.e.},
only tensile impact results in quasi-pressure shock. Panel \ref{fig:GentMQ_general}(b)
shows a unique state of initial shear ($\ishear=1.7$), at which any
axial impact will create quasi-pressure shock. As pointed out in Sec.$\ $\ref{sec:Gent},
the value of $\cpressure$ decreases as $\ishear$ increases, and
for $\ishear=1.7$ its value is 0. This implies that in a material
that was strained accordingly, any axial impact will propagate faster
than $\pressurewavespeed\left(\ipressure=0\right)$, hence coalesce
into shock.

By comparing panels \ref{fig:GentMQ_general}(d)-\ref{fig:GentMQ_general}(f)
where $\ishear$ is fixed and $\ipressure$ is increased, we observe
that $\ipressure$ has little effect on the value of $\cpressure$
and the threshold value of impact for tensile shock. The curve that
separates regions of quasi-pressure smooth waves from corresponding
shocks passes through $\ipressure$, and when $\ipressure>0$ (resp.$\ $$\ipressure<0$)
the region of impact that creates smooth quasi-pressure waves is narrower
(resp.$\ $wider).

\begin{figure}[!t]
\begin{raggedright}
\centering\sidesubfloat[]{\includegraphics[scale=0.32]{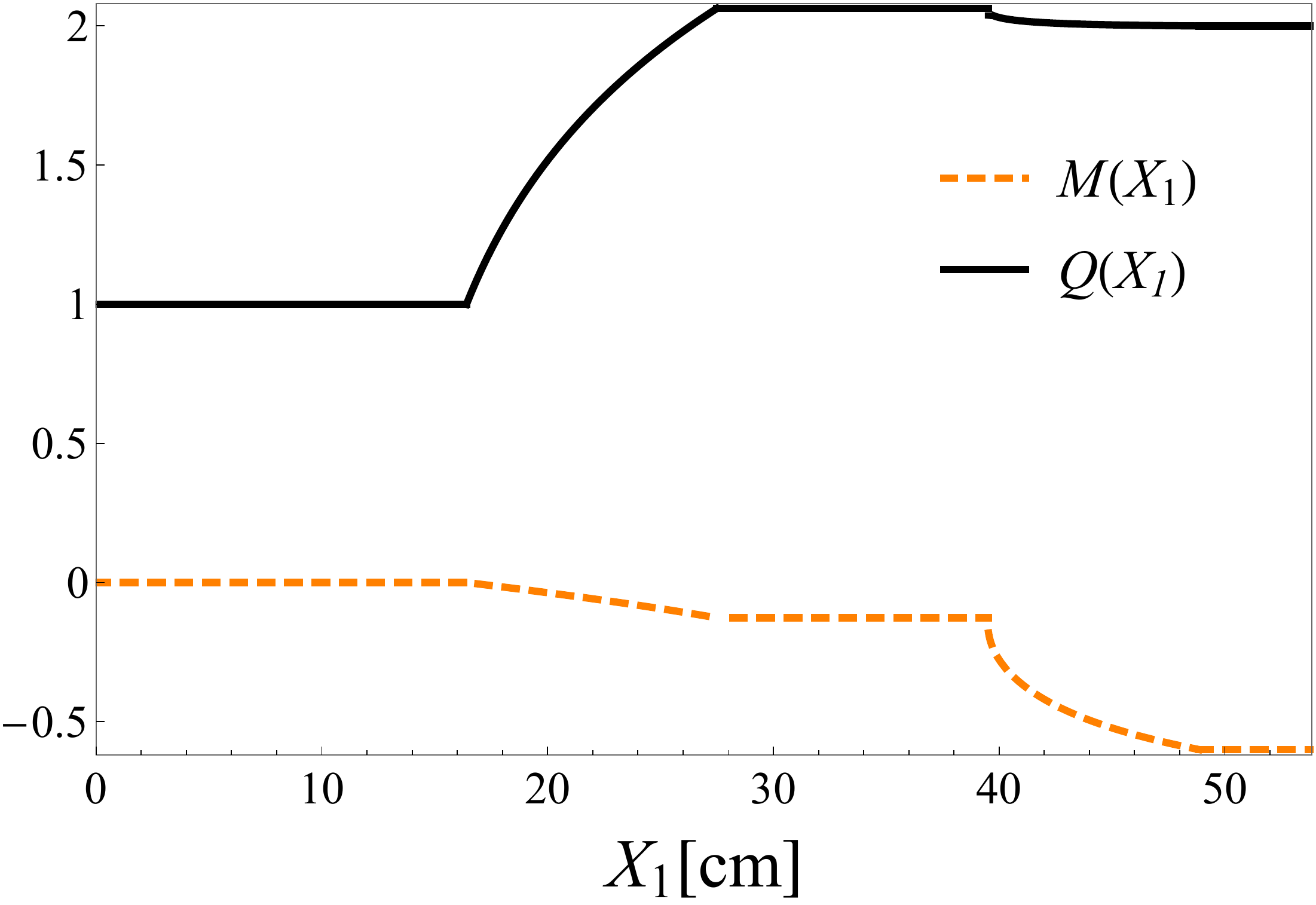}}  $\;\;\;$\sidesubfloat[]{\includegraphics[scale=0.34]{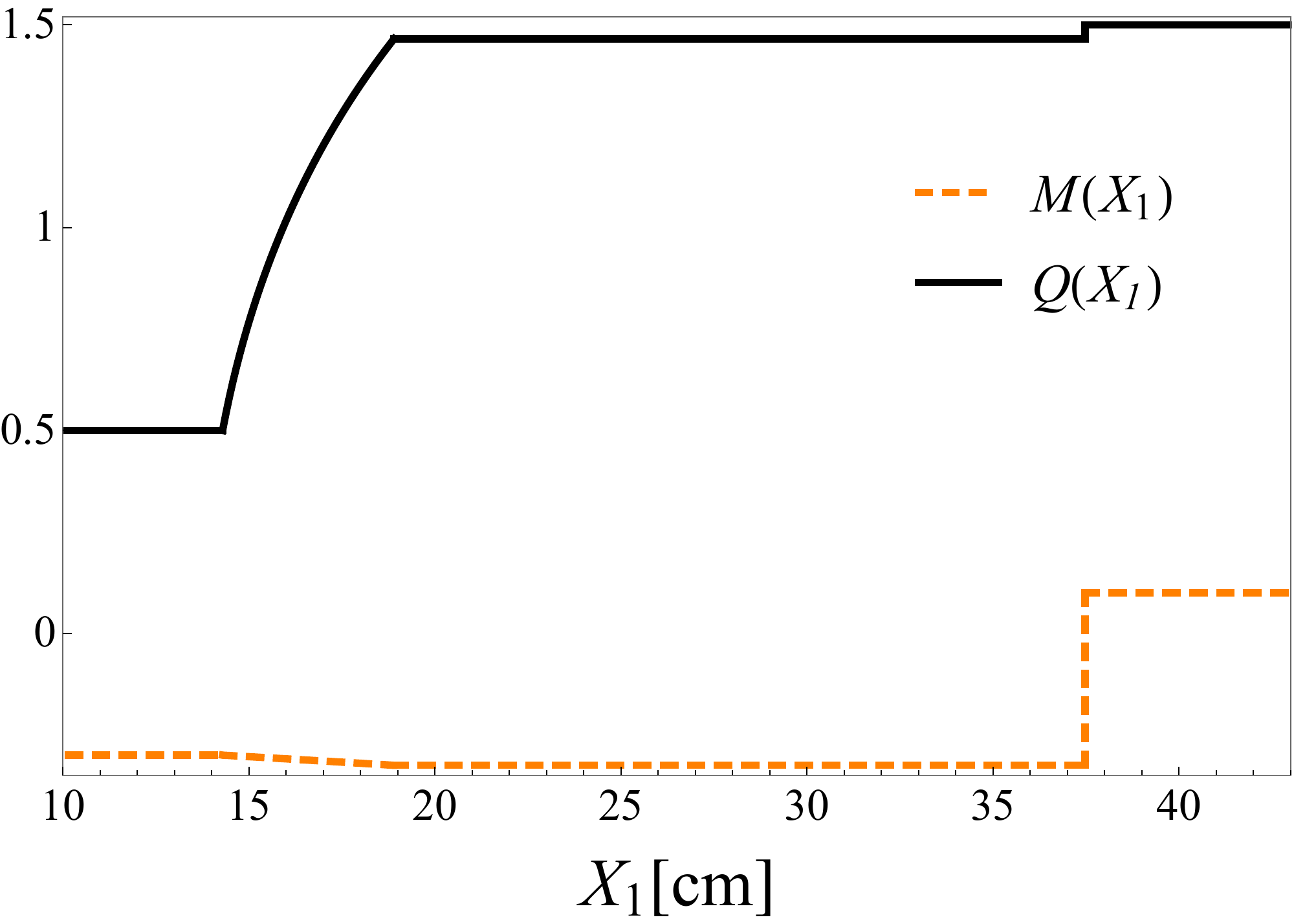}}
\par\end{raggedright}
\caption{Wave solutions for a Gent material with the parameters $\protect\bulk=1\,\mathrm{MPa},\;\protect\shear=200\,\mathrm{kPa},\;\protect\refden=1000\,\mathrm{\mathrm{kg/m^{3},\ \mathit{\protect\jm=\mathrm{10}}}}$.
Each panel shows $\protect\pressurestrain$ (solid-black) and $\protect\shearstrain$
(dashed-orange) as function of $X_{1}$, at $t=10\mathrm{ms}$, for
the conditions (a) $\protect\ipressure=-0.6,\;\protect\ishear=2,\ \protect\bpressure=0,\ \protect\bshear=1$;
(b) $\protect\ipressure=0.1,\;\protect\ishear=1.5,\ \protect\bpressure=-0.3,\ \protect\bshear=0.5$.\label{fig:MQGentExample}}
\end{figure}

We conclude our study by evaluating in Fig.$\ $\ref{fig:MQGentExample}
the strain fields $\pressurestrain$ (dashed orange) and $\shearstrain$
(solid black) from the smooth-smooth (panel \ref{fig:MQGentExample}a)
and smooth-shock (panel \ref{fig:MQGentExample}b) waves at the conditions
$\left(\ipressure,\ishear,\bpressure,\bshear\right)=\left(-0.6,2,0,1\right)$\footnote{The procedure of obtaining $\upressure$ for $\left(\ipressure,\ishear,\bpressure,\bshear\right)=\left(-0.6,2,0,1\right)$
is illustrated in Appendix \ref{Appendix-A}.} and $\left(0.1,1.5,-0.3,0.5\right)$, respectively. Evidently, quasi-shear
waves excite also axial strains, \emph{e.g.}, the interval $16\,\mathrm{cm}<X_{1}<28\,\mathrm{cm}$
in panel \ref{fig:MQGentExample}(a), and quasi-pressure shocks excite
also shear strains, \emph{e.g.}, $X_{1}=35\,\mathrm{cm}$ in panel
\ref{fig:MQGentExample}(b).

\floatsetup[figure]{style=plain,subcapbesideposition=top}

\section{\label{Conclusion}Summary}

The general aim was to study smooth and shock waves of finite amplitude
in soft materials described by the two most prominent constitutive
models, namely, the neo-Hookean and Gent models \citep{gent96rc&t}.
Our study was carried out for the plane motion of a finitely strained
semi-infinite material, in response to combined transverse and axial
impacts. Our specific objective was to characterize the effect of
the constitutive modeling, pre-deformation and impact program have
on resultant waves. 

The analysis of the neo-Hookean model provided the following observations.
We found that the resultant axial and transverse motions in neo-Hookean
materials are uncoupled. Independently of the specific boundary conditions,
smooth shear waves propagate at a constant velocity, and thus cannot
coalesce into shock. The velocity of smooth pressure waves is a monotonically
decreasing function of the axial impact, hence coalesce into shock
when the impact compresses the material, as expected. These observations
infer that the neo-Hookean model is not adequate to describe experimental
results on shear shocks and tensile-induced shocks in soft materials
\citep{Catheline2003PRL,NIEMCZURA2011442}. 

The analysis of the Gent model is more complex, exhibiting richer
results. We found that the model predicts that the resultant axial
and transverse motions are coupled such that an axial (resp.~transverse)
impact will also create transverse (resp.~axial) displacements. Contrary
to the neo-Hookean model, the Gent model predicts that with smooth
quasi-shear waves propagate faster in sheared materials, and coalesce
into shock when the prescribed transverse impact is greater than the
initial shear state. The impact release or loading of shear may form
quasi-pressure shocks owing to the coupling between the displacements,
depending on initial deformation. The Gent model further predicts
that the velocity of smooth quasi-pressure waves is a non-monotonic
function of the axial strain, and their coalescence into shock intricately
depends on the initial deformation. Notably, compressive impact may
not be sufficient to induce a quasi-pressure shock---yet it may induce
a quasi-shear shock, where tensile impact can trigger quasi-pressure
shock---and may simultaneously trigger a quasi-shear shock. In agreement
with \citet{knowles2002impact}, who tackled the tensional problem
with a kinetic approach, we find that the tensile impact must be greater
than threshold value to induce shock. We characterize the dependency
of this value on the initial deformation, and specifically find that
the threshold is lower in the presence of pre-shear. These observations
imply that the Gent model is suitable for the modeling of shocks in
soft materials. 

The interesting and more intricate aspects of finite amplitude wave
reflection and transition from free boundaries and material interfaces
are left for future studies \citep{nair1971finite,AGRAWAL2014} . 

\section*{Acknowledgments }

We acknowledge the supports of the Israel Science Foundation, funded
by the Israel Academy of Sciences and Humanities (Grant no.$\ $1912/15),
and the United States-Israel Binational Science Foundation (Grant
no.~2014358).

\appendix

\section{\label{Appendix-A}Appendix }

\begin{figure}[!t]
\includegraphics[scale=0.4]{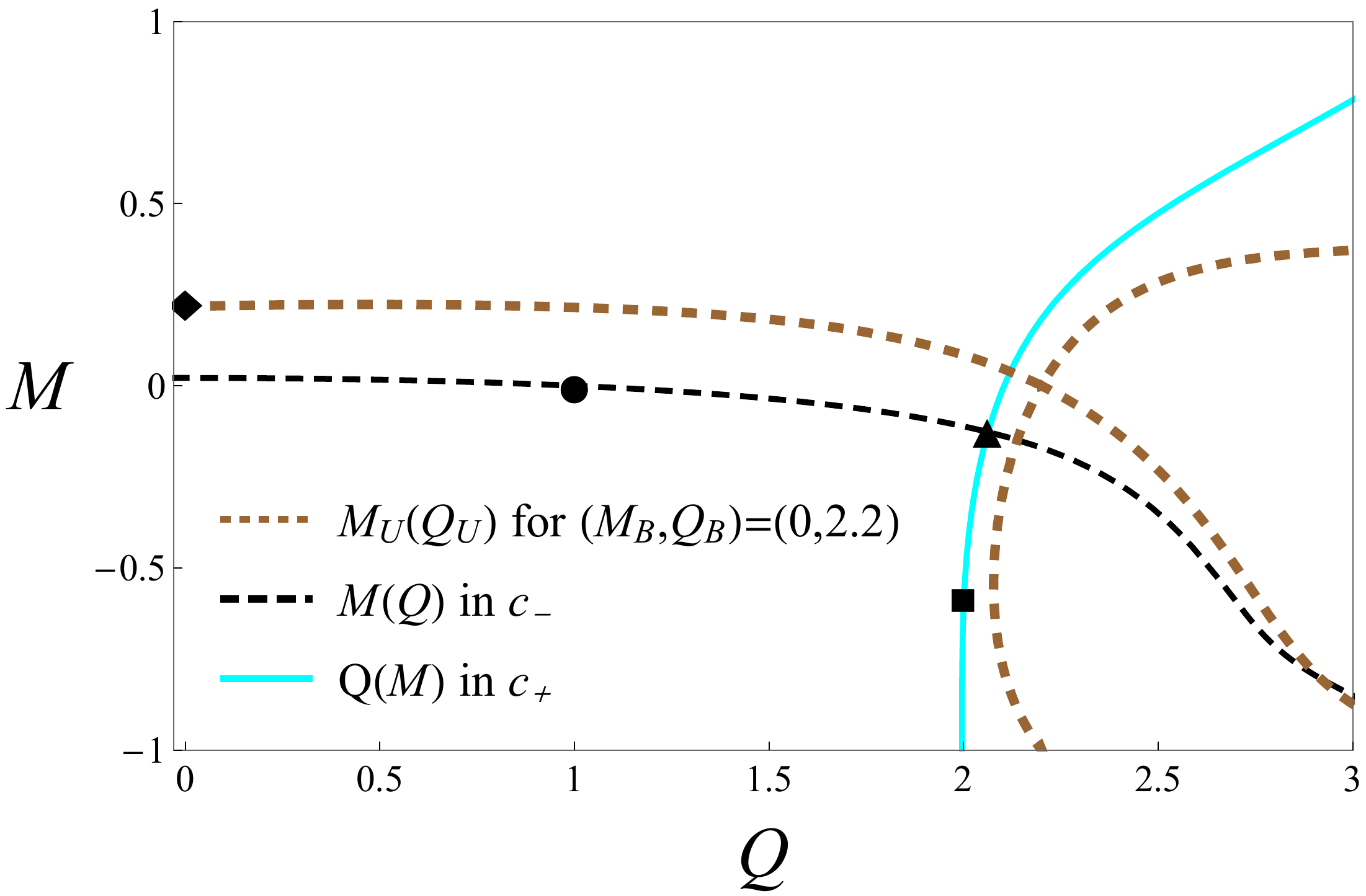}

\caption{\label{fig:PairsMQ}Pairs of ($\protect\upressure,\protect\ushear$)
that solve Eq.$\ $\eqref{eq:reducedJumpCond}$_{1}$ for the boundary
values $\left(\protect\bpressure,\protect\bshear\right)=\left(0,2.2\right)$
are denoted in brown. The value of $\protect\upressure\left(\protect\ushear=\protect\ishear=0\right)$
is denoted by the diamond mark. Solutions of Eqs.$\ $\eqref{eq:shearRegionEq}-\eqref{eq:pressureRegionEq}
are given by the dashed-black and solid-cyan, for $\left(\protect\ipressure,\protect\ishear,\protect\bpressure,\protect\bshear\right)=\left(-0.6,2,0,1\right)$.
The circle, triangle and square correspond to $\protect\bshear,\ \protect\ushear$
and $\protect\ishear$, respectively.}
\end{figure}
We graphically illustrate how we obtain the values of $\upressure$
and $\ushear$ for two of the examples in the body of the paper. Our
first example is for the conditions $\left(\ipressure,\ishear,\bpressure,\bshear\right)=\left(0,0,0,2.2\right)$.
Note that since the material is unstrained, $\pressurestrain$ and
$\shearstrain$ are uncoupled for the quasi-pressure wave. Hence,
we know that the set of equations for smooth-smooth is not compatible,
and therefore proceeded to solve the equations for shock-smooth waves.
To this end, we solve Eq.$\ $\eqref{eq:reducedJumpCond}$_{1}$ and
plot its solution in Fig.$\ $\ref{fig:PairsMQ} (dashed brown curves).
The value of $\upressure\left(\ushear=\ishear=0\right)$ is denoted
by the diamond mark.

Our second example is for $\left(\ipressure,\ishear,\bpressure,\bshear\right)=\left(-0.6,2,0,1\right)$.
To solve the equations of smooth-smooth waves, we plot in Fig.$\ $\ref{fig:PairsMQ}
the numerical solutions to Eqs.$\ $\eqref{eq:shearRegionEq}-\eqref{eq:pressureRegionEq}
for ($\pressurestrain,\shearstrain$), obtained from Wolfram Mathematica
11.3 \citet{Mathematica} We determine the values of ($\upressure,\ushear$)
from the intersection between the curves. The circle, triangle and
square correspond to $\bshear,\ \ushear$ and $\ishear$, respectively.
Furthermore, the values $\bshear$ and $\ushear$ are indicated in
Fig.$\ $\ref{fig:Gentwavespeeds}(a) by the circle and triangle marks,
respectively, showing that condition \eqref{eq:simplewaverequirement}
is satisfied for the quasi-shear wave. Similarly, the values of $\upressure$
and $\ipressure$ are indicated in Fig.$\ $\ref{fig:GentwavespeedsPRESSURE}(a)
by the as triangle and square marks, respectively, showing that condition
\eqref{eq:simplewaverequirement} is satisfied for the quasi-pressure
wave.

\global\long\def\theequation{A.\arabic{equation}}%
 \setcounter{equation}{0}

\noindent \bibliographystyle{plainnat}

\end{document}